\documentclass[journal,draftcls,onecolumn,12pt,twoside]{IEEEtran}

\usepackage{cite} 
\usepackage{amsmath,amssymb}
\usepackage{interval}
\usepackage{cool}
\usepackage{breqn}
\usepackage{wasysym}
\usepackage{dsfont}
\usepackage{graphicx}
\usepackage{times, epsfig}
\usepackage{subfig}
\usepackage{verbatim}
\newtheorem{thm}{Theorem}

\newtheorem{defn}[thm]{Definition}

\newtheorem{rk}[thm]{Remark}
\newtheorem{cor}[thm]{Corollary}

\usepackage{color}

\newlength{\figwidth}
\begin{document}

	\title{Eigenvalue Based Detection of a Signal in Colored Noise: Finite and Asymptotic Analyses}
	\author{Lahiru~D.~Chamain,~Prathapasinghe~Dharmawansa,~\IEEEmembership{Member, IEEE}, Saman~Atapattu,~\IEEEmembership{Member, IEEE}, and Chintha~Tellambura,~\IEEEmembership{Fellow, IEEE} 
		\thanks{L. D. Chamain is with the Department of Electrical and Computer Engineering, 2064 Kemper Hall, University of California Davis, 1 Shields Avenue, Davis, CA 95616 (e-mail: hdchamain@ucdavis.edu).}
		\thanks{P. Dharmawansa is with the Department of Electronic and Telecommunication Engineering, University of Moratuwa, Moratuwa 10400, Sri Lanka (e-mail: prathapa@uom.lk).}
		\thanks{S. Atapattu is with the Department of Electrical and Electronic Engineering, University of Melbourne, Parkville, VIC 3010, Australia (e-mail: saman.atapattu@unimelb.edu.au).}
	\thanks{C. Tellambura is with the Department of Electrical and Computer Engineering, University of Alberta, Edmonton, AB T6G 2R3, Canada (e-mail: ct4@ualberta.ca).}
	
}

	\maketitle
	
	

	%
    \vspace*{-10mm}
    \begin{abstract}
    Signal detection  in   colored noise with an unknown covariance matrix has a myriad of applications in diverse  scientific/engineering fields.  The test statistic is  the     largest generalized eigenvalue (l.g.e.) of  the  whitened sample covariance matrix, which is constructed via 
    $m$-dimensional $p $ signal-plus-noise samples and $m$-dimensional $n $ noise-only samples.      A  finite dimensional characterization of this statistic under the alternative hypothesis has hitherto  been an open problem.  We  answer this problem by deriving  cumulative distribution function (c.d.f.) of this l.g.e. via the   powerful orthogonal polynomial approach, exploiting  the deformed Jacobi unitary ensemble (JUE). Two special cases and an asymptotic version of the c.d.f. are also derived.   With this  new c.d.f., we comprehensively  analyze  the receiver operating characteristics (ROC) of the detector. Importantly,  when the noise-only covariant matrix is nearly rank deficient (i.e., $ m=n$), we  show that (a) when $m$ and $p$ increase such that $m/p$ is fixed, at each fixed signal-to-noise ratio (SNR), there exists an optimal ROC profile. We  also establish  a tight approximation of it;  and (b)  asymptotically,  reliable signal detection is always possible (no matter how weak the signal is) if SNR scales with $m$. 
    \end{abstract}
    
    \newpage

	\IEEEpeerreviewmaketitle

\section{Introduction}

Eigenvalue based detection of a  signal embedded in noise is a fundamental problem with a myriad of applications in diverse fields including signal processing, wireless communications, cognitive radio, bioinformatics and many  more \cite{trees2001detection,trees2002optimum,Nadakuditi2008sp,Nadakuditi2010jsacsp,Debbah2011it,Couillet2013it,Nadakuditi2013sp,Nadakuditi2017it}. Thus,  sample eigenvalue (of the sample covariance matrix) based detection has gained   prominence recently (\cite{Couillet2011book,Bianchi2011tit} and references therein). In this context, the largest sample eigenvalue based detection, also known as the Roy's largest root test \cite{Mardia1979book}, has been popular among detection theorists. Under the common  Gaussian setting with white noise, this amounts to the use of  the largest eigenvalue of a Wishart matrix having a so-called spiked covariance \cite{Baik2005annprob,Baik2006jma,Paul2007stat,Karoui2007annprob,Hoyle2007,Nadler2008}. 

However,  colored noise (or correlated noise) occurs in multitudes of applications  \cite{Maris2003biomed,Sekihara1997biomed,Sekihara1999biomed,Vinogradova2013sp,Hiltunen2015sp,Nadakuditi2017it}. In this case, we can utilize the maximum  eigenvalue  of the matrix formed by whitening the signal-plus-noise sample covariance matrix with the noise-only sample covariance matrix.   For this estimator, Nadakuditi and Silverstein \cite{Nadakuditi2010jsacsp} proposed a framework to use the generalized eigenvalues of the whitened signal-plus-noise sample covariance matrix for detection. The assumption of having the noise only sample covariance matrix is realistic in many practical situations as detailed in \cite{Nadakuditi2010jsacsp}. The  fundamental {\it high dimensional} limits of the generalized sample eigenvalue based detection in colored noise have  been thoroughly investigated in  \cite{Nadakuditi2010jsacsp}. However, to our best knowledge, a tractable {\it finite dimensional} analysis is not available in the literature. Thus,  in this paper, we characterize the statistics of the Roy's largest root in the finite dimensional colored noise setting. Moreover, we investigate certain limiting behaviors of the Roy's largest root to  deepen our understanding of the  classical detection problem in colored noise. These limiting expressions are derived based on their finite dimensional counterparts, whereas in the literature, it is customary to use entirely different tools for finite and asymptotic analyses. 

The Roy's largest root of the generalized eigenvalue detection problem in the Gaussian setting amounts to finite dimensional characterization of the largest eigenvalue of the deformed Jacobi ensemble. Various asymptotic expressions (high dimensional and high signal-to-noise ratio) for it have been derived in \cite{Johnstone2017Biometrika,Dharmawansa2014arx,Dharmawansa2014arx2,wang2017stat} for deformed Jacobi ensemble. However, finite dimensional expressions are available for Jacobi ensemble only (without  deformation) \cite{koev2005distribution,Dumitriu2012phys,Dumitriu2008siam}. Although finite dimensional, these expressions are not amenable to further manipulations. Therefore, in this paper, we present a simple and tractable closed form solution to the cumulative distribution function (c.d.f.) of the maximum eigenvalue of the deformed Jacobi ensemble. This expression further facilitates the analysis of the receiver operating characteristics (ROC) of the Roy's largest root test. All these results are made possible due to a novel alternative joint eigenvalue density function that we have derived based on the contour integral approach due to \cite{Dharmawansa2014stat,ONATSKI2014rmt,Passemier2015jmulti,Mo2012,meWang}. 

The key results developed in this paper enable us to understand the joint effect of the system dimensionality ($m$), the number of  signal-plus-noise samples  ($p$) and noise-only samples  ($n$), and the signal-to-noise ratio ($\gamma$) on the ROC. For instance, the relative disparity between $m$ and $n$ improves the ROC profile for fixed values of the other parameters. However, the general finite dimensional ROC expressions turns out to give little analytical insights. Therefore, to obtain more insights, we have particularly focused on the case for which the system dimensionality equals the number of the noise-only samples  (i.e., $m=n$). Since this equality is the minimum requirement for the validity of the  whitening operation, from the ROC perspective, it corresponds to the worst possible case when then other parameters being fixed. It turns out that, in this  scenario, when $p$ increases for fixed $m,n$ and $\gamma$, the ROC profile improves. In this respect, the ROC profile converges to a limiting profile as $p\to\infty$. In contrast, when we increase $p$ and $m$ simultaneously such that $m/p$ is a constant ($\leq 1$) for fixed $\gamma$, we can observe an optimal ROC profile for some special values of $p$ and $m$. However, as $p,m,n\to \infty$ such that $m/p$ approaches a constant ($\leq 1$) (the high dimensional limit) and $m/n=1$ for fixed $\gamma$, the maximum eigenvalue tend to lose its detection power. This phenomenon amounts to stating that the maximum eigenvalue has no power below the phase transition. This has been observed in  random matrix theory literature  \cite{Johnstone2018arx,wang2017stat,Zhigang2015stat,Nadakuditi2010jsacsp,Dykstra1978stat}.
Be that as it may, the most interesting result emerged from our analysis is that, when $\gamma$ scales with $m$ under the latter assumptions, the ROC attains a finite limit. In other words, the maximum eigenvalue still retains its detection power in the high dimension when $\gamma$ scales with $m$ as $m\to\infty$. For instance, under Rayleigh fading, as $m\to\infty$, $\gamma$ scales with $m$ (due to the strong law of large numbers). Therefore, the above insight can be of paramount importance in designing future wireless communication systems (5G and beyond). 

The remainder of this paper is organized as follows. In Section II, we formulate the classical detection problem in unknown colored noise. A new c.d.f. expression for the maximum eigenvalue (i.e., Roy's largest root) of the deformed Jacobi unitary ensemble is derived in Section III. It also gives certain particularizations of the general c.d.f. expression. Subsequently, Section IV investigates the ROC characteristics of the Roy's largest root test in the light of the c.d.f. derived in Section III. Moreover, the interplay between the system dimensionality, the number of  signal-plus-noise samples, and the noise-only samples  has been analytically characterized in Section IV. Finally, conclusions are drawn in Section V.

The following notation is used throughout this paper. The superscript $(\cdot)^\dagger$ indicates the Hermitian transpose, $\text{det}(\cdot)$ denotes the determinant of a square matrix, $\text{tr}(\cdot)$ represents the trace of a square matrix, and $\text{etr}(\cdot)$ stands for $\exp\left(\text{tr}(\cdot)\right)$. The $n\times n$ identity matrix is represented by $\mathbf{I}_n$ and the Euclidean norm of a vector $\mathbf{w}$ is denoted by $||\mathbf{w}||$. A diagonal matrix with the diagonal entries $a_1,a_2,\ldots, a_n$ is denoted by $\text{diag}(a_1, a_2,\ldots,a_n)$. We denote the $m\times m$ unitary group by $U(m)$. Finally, we use the following notation to compactly represent the
determinant of an $n\times n$ block matrix:
\begin{equation*}
\begin{split}
\det\left[a_{i}\;\; b_{i,j}\right]_{\substack{i=1,2,\ldots,n\\
 j=2,3,\ldots,n}}&=\left|\begin{array}{ccccc}
 a_{1} & b_{1,2}& b_{1,3}& \ldots & b_{1,n}\\
  a_{2} & b_{2,2}& b_{2,3}& \ldots & b_{2,n}\\
  \vdots & \vdots & \vdots &\ddots & \vdots \\
  a_{n} & b_{n,2}& b_{n,3}& \ldots & b_{n,n}
 \end{array}\right|.
 \end{split}
\end{equation*}

	\section{Problem formulation}
	
	Consider the following generic signal detection problem in colored Gaussian noise 
	\begin{align*}
	\mathbf{x}=\sqrt{\rho}\mathbf{h} s+\mathbf{n}
	\end{align*}
	where $\mathbf{x,h}\in\mathbb{C}^{m}$ are $m$-dimensional complex vectors,  $\rho>0$ is a signal power measure,  $s\sim\mathcal{CN}(0,1)$ is a complex Gaussian transmit symbol  and  $\mathbf{n}\sim \mathcal{CN}_m(\mathbf{0}, \boldsymbol{\Sigma})$ is random Complex Gaussian  noise vector with  covariance matrix $\boldsymbol{\Sigma}$, which may or may not be known at the detector.  The classical signal detection problem  amounts to the  following hypothesis testing problem: 
	\begin{align*}
	&\mathcal{H}_0:\; \rho=0\;\;\;\;\;\; \text{Signal is absent}\\
	& \mathcal{H}_1:\; \rho>0 \;\;\;\;\; \text{Signal is present}.
	\end{align*}
	Nothing that the covariance matrix of $\mathbf{x}$ can be written as
	\begin{align*}
	\mathbf{S}=\rho \mathbf{h}\mathbf{h}^\dagger+\boldsymbol{\Sigma},
	\end{align*}
	where $(\cdot)^\dagger$ denotes the conjugate transpose, we can have the following equivalent form
	\begin{align*}
	\begin{array}{ll}
	\mathcal{H}_0:\; \mathbf{R}=\boldsymbol{\Sigma} &\text{Signal is absent}\\
	\mathcal{H}_1:\; \mathbf{S}=\rho\mathbf{h}\mathbf{h}^\dagger+\boldsymbol{\Sigma} & \text{Signal is present}.
	\end{array}
	\end{align*}
 If the signal-plus-noise  covariance matrix $\mathbf  S $ and the noise covariance matrix $ \boldsymbol{\Sigma}$ were  known, we may compute matrix 
	\begin{align*}
	\boldsymbol{\Psi}=\mathbf{R}^{-1}\mathbf{S}=
	\rho \boldsymbol{\Sigma}^{-1}\mathbf{h}\mathbf{h}^\dagger+\mathbf{I}. 
	\end{align*}
	Denote the  eigenvalues of $	\boldsymbol{\Psi} $ by  $\lambda_1\leq \lambda_2\leq \ldots\leq \lambda_m$. These eigenvalues are in fact the generalized eigenvalues of the matrix pair $(\mathbf S, \mathbf R ) .$  Since the rank of $\mathbf{h}\mathbf{h}^\dagger $ is one, then $ m-1$ eigenvalues are all equal to one  ($\lambda_1=\lambda_2=\ldots=\lambda_{m-1}=1$), while the remaining  maximum eigenvalue of $\boldsymbol{\Psi}$ ($\lambda_m$) is strictly greater than one.  Thus,  the maximum eigenvalue of $\boldsymbol{\Psi}$ could be used to detect the presence of a signal \cite{Nadakuditi2010jsacsp}.  
	
	In most practical settings,  $\mathbf{R}$ and $\bf{S}$ matrices  are unknown. To circumvent this difficulty, we may  replace $\mathbf{R}$ and $\mathbf{S}$ by their sample estimates. To this end, we assume  the availability of   $p > 1 $  i.i.d.  signal-plus-noise  samples  $\{\mathbf{x}_1, \mathbf{x}_2,\ldots, \mathbf{x}_p\}$, and $n$ i.i.d.  noise-only samples  $\{\mathbf{n}_1, \mathbf{n}_2,\ldots,\mathbf{n}_n\}$. Thus,  the  sample estimates of $\mathbf{R}$ and $\mathbf{S}$ become 
	\begin{align}
	\widehat{\mathbf{R}}=\frac{1}{n}\sum_{\ell=1}^n \mathbf{n}_\ell \mathbf{n}_\ell^\dagger
	\end{align} 
	\begin{align}
	\widehat{\mathbf{S}}=\frac{1}{p}\sum_{k=1}^p \mathbf{x}_k\mathbf{x}_k^\dagger
	\end{align}
	where we assume that $n,p\geq m$ (this ensures that both $\widehat{\mathbf{R}}$ and $\widehat{\mathbf{S}}$ are positive definite with probability $1$ \cite{memuirhead2009aspects,Dykstra1978stat}). Consequently, following \cite{Nadakuditi2010jsacsp}, we form the matrix
	\begin{align}
	\widehat{\boldsymbol{\Psi}}=\widehat{\mathbf{R}}^{-1}\widehat{\mathbf{S}}
	\end{align}
	and focus on its maximum eigenvalue as the test statistic\footnote{This is also known as the Roy's largest root test which is a consequence of Roy's union intersection principle \cite{Mardia1979book}.}. As such, we have
	\begin{align*}
	&n\widehat{\mathbf{R}}\sim\mathcal{CW}_m\left(n, \boldsymbol{\Sigma}\right)\\
	& p\widehat{\mathbf{S}}\sim\mathcal{CW}_m\left(p, \boldsymbol{\Sigma}+\rho \mathbf{h}\mathbf{h}^\dagger\right)
	\end{align*}
	Noting that the eigenvalues of $ \widehat{\boldsymbol{\Psi}}$ do not change under the simultaneous transformations $\widehat{\mathbf{R}}\mapsto \boldsymbol{\Sigma}^{-1/2}\widehat{\mathbf{R}}\boldsymbol{\Sigma}^{-1/2}$, and $\widehat{\mathbf{S}}\mapsto \boldsymbol{\Sigma}^{-1/2}\widehat{\mathbf{S}}\boldsymbol{\Sigma}^{-1/2}$, without loss of generality we assume that
	$\boldsymbol{\Sigma}=\sigma^2\mathbf{I}_m$. Therefore, in what follows we focus on the maximum eigenvalue of  $ \widehat{\boldsymbol{\Psi}}$, where
	\begin{align}
	&n\widehat{\mathbf{R}}\sim\mathcal{CW}_m\left(n, \mathbf{I}_m\right)\\
	& p\widehat{\mathbf{S}}\sim\mathcal{CW}_m\left(p, \mathbf{I}_m+\gamma \mathbf{u}\mathbf{u}^\dagger\right)
	\end{align}
	with $\gamma=\rho ||\mathbf{h}||^2/\sigma^2$ and $\mathbf{u}=\mathbf{h}/||\mathbf{h}||$ being a unit vector.   
	
	Let us denote the maximum eigenvalue of $ \widehat{\boldsymbol{\Psi}}$ as $\hat{\lambda}_{\max}(\gamma)$. Now, in order to assess the performance of the maximum-eigen based  detector, we need to evaluate the detection\footnote{This is also known as the power of the test.} and false alarm probabilities. They may be expressed as
	\begin{align}
    \label{detection}
	P_D(\gamma, \mu)=\Pr\left(\hat{\lambda}_{\max}(\gamma)>\mu_{\text{th}}|\mathcal{H}_1\right)
	\end{align}
	and
	\begin{align}
    \label{alarm}
	P_F(\gamma,\mu)=\Pr\left(\hat{\lambda}_{\max}(\gamma)>\mu_{\text{th}}|\mathcal{H}_0\right)
	\end{align}
	where $\mu_{\text{th}}$ is the threshold. The $(P_D, P_F) $ pair characterizes   the detector  and is called the  ROC profile.
	
    Our main challenge  is to characterize the maximum eigenvalue of $\widehat{\boldsymbol{\Psi}}$ under the alternative $\mathcal{H}_1$. This particular matrix is also referred to as the multivariate $F$ matrix in the statistics literature  \cite{memuirhead2009aspects}. It is also related to the so called Jacobi ensemble in random matrix theory \cite{me11},\cite{forrester2010log}. The joint eigenvalue distribution of the $F$ (also Jacobi ensemble) matrix has been well documented in the literature \cite{memuirhead2009aspects}, \cite{me11}, \cite{meJames}. The extreme eigenvalues of $F$ under the null has been characterized in \cite{koev2005distribution,Dumitriu2012phys,Dumitriu2008siam}  in terms of hypergeometric function of one matrix argument. To gain more insights into the behavior of the extreme eigenvalues, focus has been shifted to various asymptotic domains (high dimensionality or  high SNR). In this respect, various asymptotic expressions for the extreme eigenvalues, under the null, have been established in \cite{Johnstone2008stat, jiang2013Bernoulli,Johnstone2017Biometrika,Dharmawansa2014arx}. Recently, capitalizing on new contour integral representations of hypergeometric functions of matrix arguments by \cite{Mo2012, Forrester2013rmt, ONATSKI2014rmt, Passemier2015jmulti, Dharmawansa2014stat}, several new asymptotic results (including phase transition phenomena) for the maximum eigenvalue, under the alternative, have been established \cite{Dharmawansa2014arx2}. Also, the authors in \cite{Nadakuditi2010jsacsp,wang2017stat,Zhigang2015stat} have employed the Stiltjes transform technique to relax the Gaussian assumption,  thereby establishing the universality nature of the above results. 
	Despite those  asymptotic results, a finite-dimensional characterization of the maximum eigenvalue under the alternative hypothesis  has been an open problem. Therefore, in this paper, we attack this problem by exploiting   orthogonal polynomial techniques due to Mehta \cite{me11} to obtain a closed-form solution.  In particular, we derive an expression which contains a determinant whose dimension depends through the relative difference between $m$ and $n$. Consequently, this property is used to establish an interesting asymptotic result on the maximum eigenvalue under the alternative hypothesis.

	\section{C.D.F. of the Maximum Eigenvalue}
    
    Before proceeding further, we present some fundamental results pertaining to the  joint eigenvalue distribution of an $F$-matrix and Jacobi polynomials.
    
    \subsection{Preliminaries}
\begin{defn}\label{F_Def}
Let $\mathbf{W}_1\sim\mathcal{W}_m\left(p,\boldsymbol{\Sigma}\right)$ and $\mathbf{W}_2\sim\mathcal{W}_m\left(n,\mathbf{I}_m\right)$ be two independent Wishart matrices with $p,n\geq m$. Then the joint eigenvalue density of the ordered eigenvalues, $\lambda_1\leq \lambda_2\leq \ldots\leq \lambda_m$, of $\mathbf{W}_1\mathbf{W}_2^{-1}$ is given by \cite{meJames}
\begin{align}
\label{james_joint}
f(\lambda_1,\lambda_2,\cdots,\lambda_m)=\frac{\mathcal{K}_1(m,n,p)}{\text{det}^p\left(\boldsymbol{\Sigma}\right)}
	\prod_{j=1}^m \lambda_j^{p-m}\Delta_m^2(\boldsymbol{\lambda})
	{}_1\widetilde F_0\left(p+n;-\boldsymbol{\Sigma}^{-1}, \boldsymbol{\Lambda}\right)
\end{align}
where  ${}_1\widetilde F_0\left(\cdot;\cdot,\cdot\right)$ is the generalized complex hypergeometric function of two matrix arguments, $\Delta_m^2(\boldsymbol{\lambda})=\prod_{1\leq i<j\leq m}\left(\lambda_j-\lambda_i\right)$ is the Vandermonde determinant, $\boldsymbol{\Lambda}=\text{diag}\left(\lambda_m,\ldots,\lambda_1\right)$, and 
\begin{align*}
\mathcal{K}_1(m,n,p)=\frac{\pi^{m(m-1)}\widetilde{\Gamma}_m(n+p)}{\widetilde{\Gamma}_m(m)\widetilde{\Gamma}_m(n)\widetilde{\Gamma}_m(p)}
\end{align*}
with the complex multivariate gamma function is written in terms of the classical gamma function $\Gamma(\cdot)$ as
\begin{align*}
\widetilde{\Gamma}_m(n)=\pi^{\frac{1}{2}m(m-1)} \prod_{j=1}^m \Gamma\left(n-j+1\right).
\end{align*}
\end{defn}

\begin{defn}
Jacobi polynomials can be defined as follows \cite[eq. 5.112]{me12}
	\begin{equation}\label{jacobidef1}
	P_{n}^{(a,b)}(x) = \sum_{k=0}^{n}\binom{n+a}{n-k}\binom{n+k+a+b}{k}\left(\frac{x-1}{2}\right)^{k} \hspace{6mm} \text{for } a,b>-1
	\end{equation} 
    
    where $\binom{n}{k}=\frac{n!}{(n-k)! k!}$ with $n\geq k\geq 0$.
    \end{defn}
    We may alternatively express
	the Jacobi polynomial as \cite{me12}
	\begin{equation}\label{jacobidef2}
P_{n}^{(a,b)}(x) = \binom{n+a}{a}\Hypergeometric{2}{1}{-n,n+a+b+1}{1+a}{\frac{1-x}{2}}
	\end{equation}
    where ${}_2F_1(\cdot;\cdot;\cdot)$ is the Gauss hypergeometric function. Following (\ref{jacobidef2}), the successive derivatives of the Jacobi polynomial can be written as
	\begin{equation}\label{jacobiDerivative}
	\frac{{\rm d}^{k}}{{\rm d}x^{k}}P_{n}^{(a,b)}(x) = 2^{-k}(n+a+b+1)_{k}P_{n-k}^{(a+k,b+k)}(x)
	\end{equation} 
    where $(a)_k=a(a+1)\ldots(a+k-1)$ with $(a)_0=1$ denotes the Pochhammer symbol. It is noteworthy that, for a negative integer $-n$ with $n\in \mathbb{Z}^+$, we have \cite{me12}
    \begin{align*}
    (-n)_k=\left\{\begin{array}{ll}
	\frac{(-1)^{k}n!}{(n-k)!} &  \text{if } 0\leq k\leq n\\
	0 & \text{if } k> n.
    \end{array}\right.
    \end{align*}
\subsection{Finite Dimensional Analysis of the C.D.F.}
Armed with these  preliminary definitions, now we focus on deriving the new c.d.f. for the maximum eigenvalue of $\mathbf{W}_1\mathbf{W}_2^{-1}$ when the covaraince matrix $\boldsymbol{\Sigma}$ takes the so called rank-$1$ spiked form. That is,  the covariance matrix can be decomposed as
\begin{align}
\label{spike_cov}
\boldsymbol{\Sigma}=\mathbf{I}_m+\eta \mathbf{vv}^\dagger=\mathbf{V}\text{diag}\left(1+\eta, 1, 1,\ldots, 1\right)\mathbf{V}^\dagger 
\end{align}
where $\mathbf{V}=\left(\mathbf{v}\; \mathbf{v}_2\; \ldots \mathbf{v}_m\right)\in\mathbb{C}^{m\times m}$ is a unitary matrix and  $\eta\geq 0$.  Before developing our method, it is important to highlight the difficulty of a direct solution via (\ref{james_joint}).  Following Khatri \cite{Khatri}, the hypergeometric function of two matrix arguments given in the join density (\ref{james_joint}) can be written as a ratio between the determinants of two $m\times m$ square matrices. Since the eigenvalues of the matrix $\boldsymbol{\Sigma}^{-1}$ are such that $1/(1+\eta)$ has algebraic multiplicity one  and $1$ has algebraic multiplicity $m-1$, the resultant  ratio takes an indeterminate form. Therefore, one has to repeatedly apply  L'Hospital's  rule to obtain a deterministic expression. However, the resulting  expression is not amenable to apply Mehta's \cite{me11} orthogonal polynomial technique. Therefore, to  apply it, we first  derive an alternative joint eigenvalue density expression.  This alternative derivation technique has also been used earlier in \cite{Dharmawansa2014stat} to derive a single contour integral representation for the joint eigenvalue density when the matrices are real\footnote{However,  when the matrices are real, the hypergeometric function of two matrix arguments does not admit such a determinant representation.}. The following corollary gives the alternative joint density  expression.  

 \begin{cor}\label{joint_eig_pdf}
 Let $\mathbf{W}_1\sim\mathcal{W}_m(p,\mathbf{I}_m+\eta \mathbf{v}\mathbf{v}^\dagger)$ and $\mathbf{W}_2\sim\mathcal{W}_m(n,\mathbf{I}_m)$ be independent Wishart matrices with $m\leq p,n$ and $\eta\geq 0$. Then the joint density of the ordered eigenvalues $0\leq \lambda_1\leq \lambda_2\leq\cdots\leq \lambda_m<\infty$ of $\mathbf{W}_1\mathbf{W}_2^{-1}$ is given by
		\begin{align}
		\label{jpdf}
		f(\lambda_1,\lambda_2,\cdots,\lambda_m)=
		f_{\text{uc}}(\lambda_1,\lambda_2,\cdots,\lambda_m) f_{\text{cor}}(\lambda_1,\lambda_2,\cdots,\lambda_m)
		\end{align}
		where
		\begin{align}
		f_{\text{uc}}(\lambda_1,\lambda_2,\cdots,\lambda_m)=\mathcal{K}_1(m,n,p)
		\prod_{j=1}^m \frac{\lambda_j^{p-m}}{(1+\lambda_j)^{p+n}} \Delta_m^2(\boldsymbol{\lambda}),
		\end{align}
		\begin{align*}
		f_{\text{cor}}(\lambda_1,\lambda_2,\cdots,\lambda_m)=\frac{\mathcal{K}_2(m,n,p)}{\eta^{m-1}(1+\eta)^{p+1-m}}
		\prod_{j=1}^m (1+\lambda_j)
		\sum_{k=1}^m
		\frac{(1+\lambda_k)^{p+n-1}}{\displaystyle\prod_{\substack{j=1\\
					j\neq k}}^m(\lambda_k-\lambda_j)
		\left(1+\frac{\lambda_k}{\eta+1}\right)^{p+n+1-m}},
		\end{align*}
        and 
		\begin{equation*}
		\mathcal{K}_2(m,n,p)=\frac{(m-1)!(p+n-m)!}{(p+n-1)!},
		\end{equation*}
 \end{cor}
{\it Proof}: See Appendix \ref{appPrathapa}.
\begin{rk}
		It is worth noting that the function $f_{\text{uc}}(\lambda_1,\lambda_2,\cdots,\lambda_m)$ denotes the joint density of the ordered eigenvalues of $\mathbf{W}_1\mathbf{W}_2^{-1}$ corresponding to the case $\mathbf{W}_1\sim\mathcal{W}_m(p,\mathbf{I}_m)$ and $\mathbf{W}_2\sim\mathcal{W}_m(n,\mathbf{I}_m)$.
	\end{rk}

	To facilitates further analysis, nothing that the continuous mapping $h:x\mapsto\frac{x}{x+1},\;x\geq 0$ is strictly increasing (i.e., order preserving), we use the variable transformations
	\begin{equation}
	x_j=\frac{\lambda_j}{1+\lambda_j},\;\; j=1,2,\cdots,m,
	\end{equation}
	with $0\leq x_1\leq x_2\leq\cdots\leq x_m<1$ in (\ref{jpdf}) to obtain
	\begin{align}
	\label{jpdfalt}
	g(x_1,x_2,\cdots,x_m)=\frac{\mathcal{K}_3(m,n,p)}{\eta^{m-1}(1+\eta)^{p+1-m}} & \Delta_m^2(\mathbf{x})
	\displaystyle \prod_{j=1}^mx_j^{p-m}(1-x_j)^{n-m}\nonumber\\
	& \times \displaystyle \sum_{k=1}^m
	\frac{1}{\displaystyle \prod_{\substack{j=1\\
				j\neq k}}^m(x_k-x_j)
	\left(1-\displaystyle \frac{\eta}{\eta+1}x_k\right)^{p+n+1-m}}
	\end{align}
	where $\mathcal{K}_3(m,n,p)=\mathcal{K}_1(m,n,p)\mathcal{K}_2(m,n,p)$.
    
    The  joint eigenvalue density (\ref{jpdfalt})  in turn facilitates the use of Mehta's orthogonal polynomial approach in our subsequent c.d.f. analysis. 
	\begin{rk}
		Alternatively, (\ref{jpdfalt}) represents the joint density of the ordered eigenvalues of {\it deformed Jacobi ensemble}, $\mathbf{W}_1(\mathbf{W}_2+\mathbf{W}_1)^{-1}$ with $\mathbf{W}_1\sim\mathcal{W}_m(p,\mathbf{I}_m+\eta \mathbf{vv}^\dagger)$ and $\mathbf{W}_2\sim\mathcal{W}_m(n,\mathbf{I}_m)$.
	\end{rk}
	
   We now consider the main contribution of  of this paper, namely, the  derivation of the c.d.f. of the maximum eigenvalue. By the definition, the c.d.f. of $ x_{\max} $ (i.e., $x_m$) can be written as,
	\begin{align}
    \label{cdfdef}
	F_{x_{\max}}(t)=\Pr(x_{\max}\leq t)=
	 \int_{0\leq x_{1}\leq x_{2} \leq \cdots\leq x_{m}\leq t}g(x_1,x_2,\cdots,x_m) \text{ d}\textbf{x}
	\end{align}
    where, for notational concision, we have used ${\rm d}\mathbf{x}={\rm d}x_1{\rm d}x_2\ldots {\rm d}x_m$.
By evaluating the above Selberg-type integral, the c.d.f. of $ x_\max $ can be found and hence the c.d.f. of $ \lambda_\max $, which is given by the the following theorem.
	\begin{thm}\label{alter}
		Let $\mathbf{W}_1\sim\mathcal{W}_m(p,\mathbf{I}_m+\eta \mathbf{vv}^\dagger)$ and $\mathbf{W}_2\sim\mathcal{W}_m(n,\mathbf{I}_m)$ be independent with $m\leq p,n$ and $\eta\geq 0$. Then the c.d.f. of the maximum eigenvalue $ \lambda_{\max} $ of $\mathbf{W}_1\mathbf{W}_2^{-1}$ is given by
		\begin{dmath}\label{cdfthm}
			F^{(\alpha)}_{\lambda_{\max}}(t;\eta)
			=\dfrac{\mathcal{K}(m,p,\alpha)}{(p-1)!(1+\eta)^{p}}\left(\dfrac{t}{1+t}\right)^{m(\alpha+\beta+m)}
			\det\left[\Phi_{i}(t,\eta)\hspace{3mm} \Psi_{i,j}(t)\right]_{\substack{i=1,2,...,\alpha+1\\j=2,3,...,\alpha+1}}
		\end{dmath}
		where
        \begin{dmath*}
			\Psi_{i,j}(t)= (m+i+\beta-1)_{j-2}P_{m+i-j}^{(j-2,\beta+j-2)}\left(\frac{2}{t}+1\right),
		\end{dmath*}
		\begin{dmath*}
		\Phi_{i}(t,\eta)=\mathcal{Q}_i(m,n,p)\sum_{k=0}^{\alpha-i+1}\frac{(p+i-1)_k(\alpha-i+2)!}{k!(p+m+2i-2)_k(\alpha-i-k+1)!} \frac{\left(\eta t\right)^{k+i-1}\left((1+\eta)(1+t)\right)^{p+k}}{\left(1+\eta +t\right)^{p+k+i-1}},
		\end{dmath*}
		\begin{dmath*}
			\mathcal{Q}_i(m,n,p)=\frac{(n+p+i-2)!(p+i-2)!}{(p+m+2i-3)!},
		\end{dmath*}
        and 
		\begin{align*}
        \mathcal{K}(m,p,\alpha)=\prod_{j=0}^{\alpha-1}\dfrac{(p+m+j-1)!}{(p+m+2j)!}
        \end{align*}
		with $ \alpha = n-m$ and $\beta=p-m$.	
	\end{thm}
	{\bf Proof:} See Appendix \ref{appa}.
	\begin{rk}\label{remarkHyper}
		Alternatively, $ \Phi_{\text{i}}(t,\eta) $ can be expressed in terms of Gauss hypergeometric function as follows
		\begin{align}\label{eqHyper}
			\Phi_{i}(t,\eta)&=\mathcal{Q}_i(m,n,p)\left(\frac{\eta t}{(1+\eta)(1+t)}\right)^{i-1}\nonumber\\
            &\qquad \quad  \times \Hypergeometric{2}{1}{\beta+m+i-1,n+p+i-1}{\beta+2m+2i-2}{\frac{\eta t}{(1+\eta)(1+t)}}.
		\end{align}
	\end{rk}
	
  The new exact c.d.f. expression for the maximum eigenvalue of $\mathbf{W}_1\mathbf{W}_2^{-1}$, which contains the determinant of a square matrix whose dimension depends on the difference $\alpha=n-m$, is highly desirable when the difference between $m$ and $n$ is small irrespective  of their individual magnitudes. For instance, when $n=m$ ($\alpha=0$) the determinant vanishes and we obtain a scalar result. This concise result is one of the many advantages of using the orthogonal polynomial approach. This key representation, also facilitates the derivation of the limiting  distribution of the maximum eigenvalue (when $m,n\to\infty$ such that $m-n$ is fixed).
  
For some special values of $\alpha$ and $\eta$,  the c.d.f. expression (\ref{cdfthm}) admits the following simple forms.
\begin{cor}\label{nullcdf}
The exact c.d.f. of the maximum eigenvalue of $\mathbf{W}_1\mathbf{W}_2^{-1}$ when  $\eta = 0$ is given by
		\begin{dmath}\label{final_eta_0}
			F^{(\alpha)}_{\lambda_{\max}}(t;0)
			=\mathcal{K}(m,p,\alpha)\dfrac{(n+p-1)!}{(m+p-1)!}\left(\dfrac{t}{1+t}\right)^{m(\alpha+\beta+m)}\det\left[\Psi_{i+1,j+1}(t)\right]_{i,j=1,2,...,\alpha}.
		\end{dmath}
\end{cor}
  {\it Proof}: Following (\ref{eqHyper}), it is easy to see that, when $\eta=0$, all the elements in the first column of the determinant in (\ref{cdfthm}) become zero except the first entry which is $(p-1)! (n+p-1)!/(m+p-1)!$. Therefore, we expand the determinant with its first column and shift the indices $i$ and $j$ to conclude the proof. 
  
Alternative expressions for c.d.f and p.d.f. of $x_{\max}$ ($x_\max=\lambda_\max/(1+\lambda_{\max})$) in the same scenario ($\eta=0$) are given in \cite{koev2005distribution} and \cite{Dumitriu2012phys}, respectively. However, these results are fundamentally structurally different from our expression (\ref{final_eta_0}), since they contain complex hypergeometric functions of one matrix argument. In particular, the matrix argument in \cite{koev2005distribution} assumes the form $t\mathbf{I}_m$, whereas the matrix argument in \cite{Dumitriu2012phys} takes the form $t\mathbf{I}_{\alpha-1}$. Further simplification of these expressions requires the repeated application of L'Hospital's    rule followed by the evaluation of the resultant determinants,  a  cumbersome process. In contrast,   the c.d.f. expression  (\ref{final_eta_0})  does not suffer from these  drawbacks.
  
\begin{cor}
		The exact c.d.f. of the maximum eigenvalue of $\mathbf{W}_1\mathbf{W}_2^{-1}$ when  $\alpha = 0$ is given by ( $ t \geq 0 $) 
		\begin{dmath}\label{cdfalpha0}
			F^{(0)}_{\lambda_{\max}}(t;\eta)
			=\dfrac{\left(\dfrac{t}{1+t}\right)^{mp}}{\left(1+\dfrac{\eta }{1+t}\right)^{p}} .
		\end{dmath}
	\end{cor}  
    {\it Proof}: When $\alpha=0$, the determinant in (\ref{cdfthm}) reduces to a single term given by
    \begin{align*}
   \Phi_1(t,\eta)= (p-1)!{}_2F_1\left(p,n+p;n+p;\frac{\eta t}{(1+\eta)(1+t)}\right).
    \end{align*}
    Nothing that ${}_2F_1(a,b;b;z)={}_1F_0(a;z)=(1-z)^{-a}$ with some algebraic manipulations concludes the proof.
    
    In the sequel, this  remarkably simple result (\ref{cdfalpha0}) is used to establish an important high dimensional limit for the maximum eigenvalue. Also, we have, for $\eta_2>\eta_1> 0$,
    \begin{align*}
    F^{(0)}(t;\eta_2)< F^{(0)}(t;\eta_1)< F^{(0)}(t;0).
    \end{align*}
    
    Having established the  finite dimensional c.d.f. results, we  now focus on the asymptotic characterization of the maximum eigenvalue. 
    
   \subsection{Asymptotic Analysis of the C.D.F.} 
   
   Here we characterize the asymptotic behavior of the maximum eigenvalue of $\mathbf{W}_1\mathbf{W}_2^{-1}$ by deriving various limiting c.d.f. expression for (\ref{cdfthm}). In particular, we focus on suitably centerd and scaled maximum eigenvalue in the following two important scenarios:
   \begin{enumerate}
   \item As $m,n,p\to \infty$ such that $\alpha, \beta$, and $\eta$ are fixed,
  
   \item As $m,n,p,\eta \to \infty $ such that $\frac{m}{n}\to 1$, $\frac{m}{p}\to c\in(0,1]$, and $\frac{\eta}{m}\to \theta\geq 0$.
    \end{enumerate}
    
 Asymptotic behavior of the Jacobi ensemble has been thoroughly studied in the literature (\cite{Johnstone2008stat}, \cite{forrester2010log}, \cite{jiang2013Bernoulli} and references therein). For instance,   Johnstone \cite{Johnstone2008stat} has shown that, for a large class of Jacobi ensembles, properly centered and scaled maximum eigenvalue (the high dimensional limit) admits a  Tracy-Widom type limiting distribution. Recently, Ioana \cite{Dumitriu2012phys} has derived a  new limiting p.d.f. expression for the maximum and minimum eigenvalues of the Jacobi ensemble for certain new asymptotic regimes. Despite the differences in the asymptotic regimes of their choice, one common features of all the above mentioned investigations  is that  $\mathbf{W}_1$ and $\mathbf{W}_2$ are  white Wishart matrices. In contrast, more recently, high dimensional limit of the maximum eigenvalue (including the so called universality) has been established when $\mathbf{W}_2$ has certain spiked covariance structures (akin to the structure given in (\ref{spike_cov}))  \cite{Dharmawansa2014arx2}, \cite{Nadakuditi2010jsacsp}, \cite{Johnstone2018arx}, \cite{Zhigang2015stat}, \cite{wang2017stat}. Most importantly those authors have observed a so called phase transition (also known as BBP phase transition) phenomena associated with the maximum eigenvalue. In a nutshell, phase transition means, in the high dimensional limit, when $\eta$ is below a certain critical threshold, the maximum eigenvalue does not separate from the rest of the eigenvalues\footnote{To be precise, it converges almost surely to the upper support of the limiting spectral density \cite{Nadakuditi2010jsacsp}, \cite{Zhigang2015stat}, \cite{Dharmawansa2014arx2}}, whereas when $\eta$ is above the threshold, it separates from the rest of the eigenvalues\footnote{It converges almost surely to a location above the upper support of the limiting spectral density \cite{Dharmawansa2014arx2}, \cite{Zhigang2015stat}.}. Despite all these efforts, the behavior of the maximum eigenvalue in the above two asymptotic regimes have not been addressed in the literature. Therefore, in what follows we give limiting c.d.f. expressions pertaining to the above two scenarios.
 
    \begin{thm}
    \label{thmAsymp}
		As $ m $, $ p $ and $ n $ tend to $ \infty $ such that $ \alpha=m-n$, $ \beta = p-m $, and $\eta$  are fixed, the centered and scaled maximum eigenvalue $ \displaystyle (1+\lambda_\max)/m^2 $ converges in distribution to a random variable $ X $ with
		the c.d.f. $ F^{(\alpha)}_{X}(x;\eta) $. In particular, we have
		\begin{align}
        \label{thmasylambdamax}
			\lim_{m\to\infty} F^{(\alpha)}_{\frac{1+\lambda_\max}{m^2}}(x)=F^{(\alpha)}_{X}(x)
			=\exp\left({-\frac{1}{x}}\right)\det\left[\mathcal{I}_{j-i}\left(\frac{2}{\sqrt{x}}\right)\right]_{i,j=1,2,\cdots,\alpha}
		\end{align}
		where
		$ \mathcal{I}_{k}(z)$ is the $k$-th order modified Bessel function of the first kind.
	\end{thm}
	{\bf Proof:} See Appendix \ref{appb}.
    
It is interesting to see that the limiting c.d.f. is independent of $\eta$. Due to this independence,  (\ref{thmasylambdamax})  should be the limiting c.d.f. for  $\eta=0$ as well. However, an alternative expression for the limiting p.d.f. of $x_{\max}$ when  $\eta=0$ has been given in \cite{Dumitriu2012phys}. That particular expression contains a hypergeometric function of one matrix argument, and therefore does not admit a simple form. In contrast, the  limiting c.d.f. (\ref{thmasylambdamax})  is simple from the representation as well as numerical evaluation perspectives. Since  (\ref{thmasylambdamax})  has the same form under both hypotheses, the maximum eigenvalue based test  does not have power in this particular regime. 

The following theorem characterizes the maximum eigenvalue in one of the most important high dimensional setting outlined in the above second scenario.
\begin{thm}
\label{doublepower}
As $ m $, $ p $, $ n $, and $\eta$ tend to $ \infty $ such that $m/n\to 1$, $m/p\to c\in(0,1]$, and $\eta/m\to \theta \geq 0$, the centered and scaled maximum eigenvalue $ \displaystyle (1+\lambda_\max)/m^2 $ converges in distribution to a random variable $ X $ with the c.d.f. $F_{X}(x;c,\theta)$. In particular, we have
\begin{align*}
\lim_{m\to\infty} F^{(0)}_{\frac{1+\lambda_\max}{m^2}}(x;\theta m)=F_{X}(x;c,\theta)=\exp\left({-\frac{1+\theta}{cx}}\right).
\end{align*}
\end{thm}
{\it Proof}: Following (\ref{cdfalpha0}), we take $\alpha=0$ and $p=m/c$ to yield
\begin{align*}
F^{(0)}_{\lambda_{\max}}(x;\eta)
			=\dfrac{\left(\dfrac{x}{1+x}\right)^{m^2/c}}{\left(1+\dfrac{\eta }{1+x}\right)^{m/c}},
\end{align*}
from which we obtain, noting that $\eta=\theta m$,
\begin{align}
\lim_{m\to\infty} F^{(0)}_{\frac{1+\lambda_\max}{m^2}}(x;\theta m)=F_{X}(x;\theta,c)=
\lim_{m\to\infty}\frac{\displaystyle \left(1-\frac{1}{x m^2}\right)^{m^2/c}}{\displaystyle\left(1+\frac{\theta}{xm}\right)^{m/c}}.
\end{align}
The final result now follows by evaluating the limits as $m\to \infty$.

This remarkably simple limiting c.d.f. sheds  some new light on the behavior of the maximum eigenvalue in this particular asymptotic domain. Following  \cite{Wachter1978prob}, \cite{Nadakuditi2010jsacsp}, we can easily show that, for $m/n\to 1$ and $m/p\to c\in(0,1]$, the upper support of the limiting spectral density diverges to infinity\footnote{Following \cite{Wachter1978prob}, \cite{Silverstein1985siam} we can show that the exact limiting spectral density takes the form $\frac{\sqrt{x-a}}{\pi x(x+c)}$, where $a=(1-c)^2/4\leq x<\infty $.} for fixed $\eta$. Therefore, under this scaling, the operatinal regime is below  below the phase transition, where the maximum eigenvalue has {\it no detection power}  \cite{Dharmawansa2014arx2}, \cite{Nadakuditi2010jsacsp}. In contrast, when $\eta$ also scales with $m$, it turns out that (see next section), the maximum eigenvalue {\it has  detection power} as shown in Theorem \ref{doublepower}. The reason is that the all earlier results treated $\eta$ as a constant when dealing with the high dimensional limits. This new simple result shows that, when $n,p$ and $\eta$ scale with $m$, an interesting new phenomenon  occurs.

 Having armed with the finite and asymptotic characteristics of the maximum eigenvalue of $\mathbf{W}_1\mathbf{W}_2^{-1}$, we next focus on the ROC curve  of the maximum eigenvalue based detector.

    \section{ROC of the Maximum Eigenvalue of $\widehat{\bf{\Psi}}$}
    We now investigate the behavior of detection and false alarm probabilities of  the maximum eigenvalue based test. To this end, noting that the eigenvalues of $\widehat{\boldsymbol{\Psi}}$ and $\mathbf{W}_1\mathbf{W}_2^{-1}$ are related by $\hat{\lambda}_j=(n/p)\lambda_j$, for $j=1,2,\ldots,m$, we  represent the c.d.f. of the maximum eigenvalue corresponding to $\widehat{\boldsymbol{\Psi}}$ as
    $F_{\lambda_\max}^{(\alpha)}(\kappa x;\gamma)$, where $\kappa=p/n$. For  convenient  presentation,  we treat the finite dimensional and asymptotic behaviors of the ROC in two separate sub sections.

    \subsection{Finite Dimensional Analysis}
    We first  consider the case where matrix dimensions ($m,n$, and $p$) are finite.
    Now following Theorem \ref{alter} and Corollary \ref{nullcdf} along with with (\ref{detection}), (\ref{alarm}), the detection and false alarm probabilities can be written, respectively, as
    \begin{align}
    P_D(\gamma, \mu_{\text{th}})&=1-F_{\lambda_\max}^{(\alpha)}(\kappa\mu_{\text{th}};\gamma)\\
    P_F(\mu_{\text{th}})&=1-F_{\lambda_\max}^{(\alpha)}(\kappa\mu_{\text{th}};0). 
    \end{align}
    In general, deriving a functional relationship between $P_D$ and $P_F$ by eliminating the parametric dependency on $\mu_{\text{th}}$ is challenging. However, when $\alpha$ admits zero,  an explicit relationship between them  is specified in  Corollary \ref{corasybalanced}.
    \begin{cor}
    \label{corasybalanced}
    For notational brevity, we  suppress the parameters $\gamma$ and  $\mu_{\text{th}}$ and represent the detection and false alarm probabilities, simply  as $P_D$ and $P_F$. Then, when $\alpha=0$, $P_D$ and $P_F$ are functionally related as  
    \begin{align}
    \label{rocbalanced}
    P_D=1-\frac{1-P_F}{\left(1+\gamma-\gamma\left[1-P_F\right]^{1/mp}\right)^p}.
    \end{align}
    \end{cor}



From \eqref{rocbalanced},  taken $P_D$ as a function of $\gamma$, we can easily see that, for $\gamma_1>\gamma_2$,
\begin{align*}
P_D(\gamma_2)>P_D(\gamma_1).
\end{align*}
This confirms the common observation that the SNR is positively correlated with the detection probability for a fixed value of $P_F$.

\begin{figure}
	\centering
	\subfloat[$P_D$ vs $\gamma$ for different values of $P_F$.]{
		\label{ROCscene1}
		\includegraphics[width=0.5\textwidth]{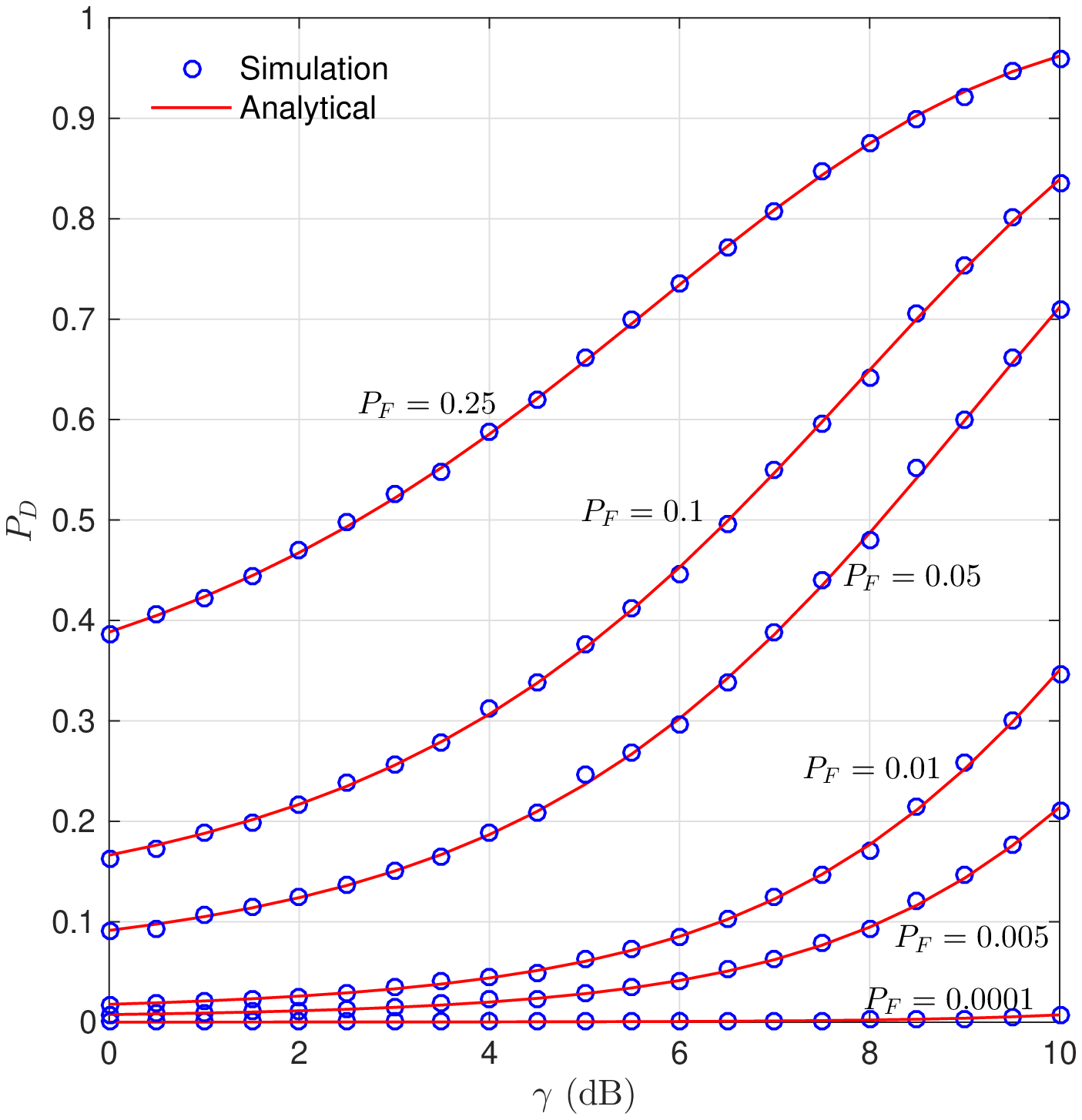}} 
	\subfloat[$P_D$ vs $P_F$ for different valued of $\gamma$.]{
		\label{ROCscene2}
		\includegraphics[width=0.5\textwidth]{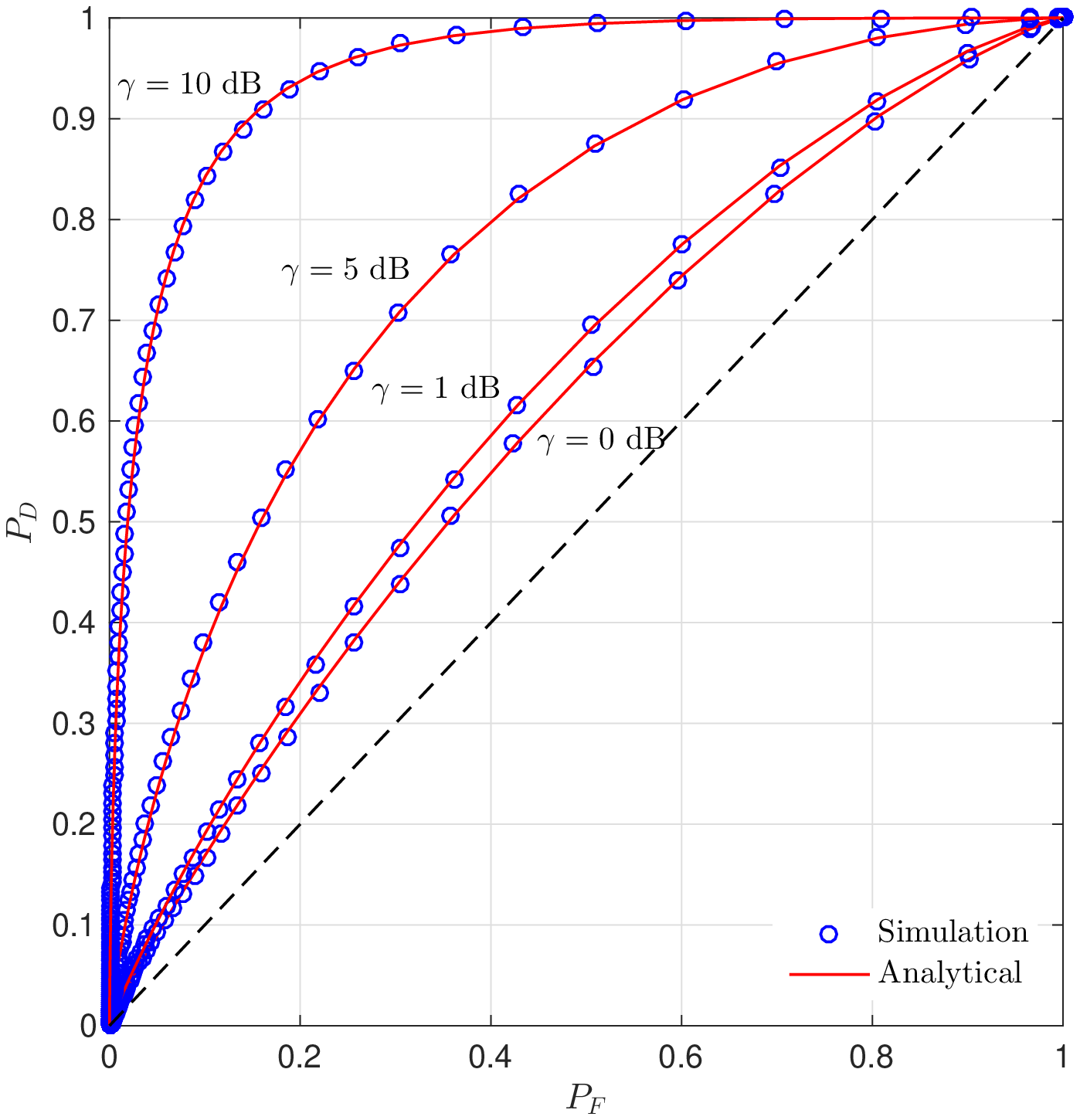}}        
	\caption{Probability of detection, $P_D$, as a function of $\gamma$ and $P_F$ for $(m,n,p)=(5,8,10)$.}
	\label{ROCscene12}
\end{figure}

    

\begin{figure}
	\centering
	\subfloat[For different $m/n$ values with $m/p=1$ and $n=10$.]{
		\label{ROCscene3}
		\includegraphics[width=0.5\textwidth]{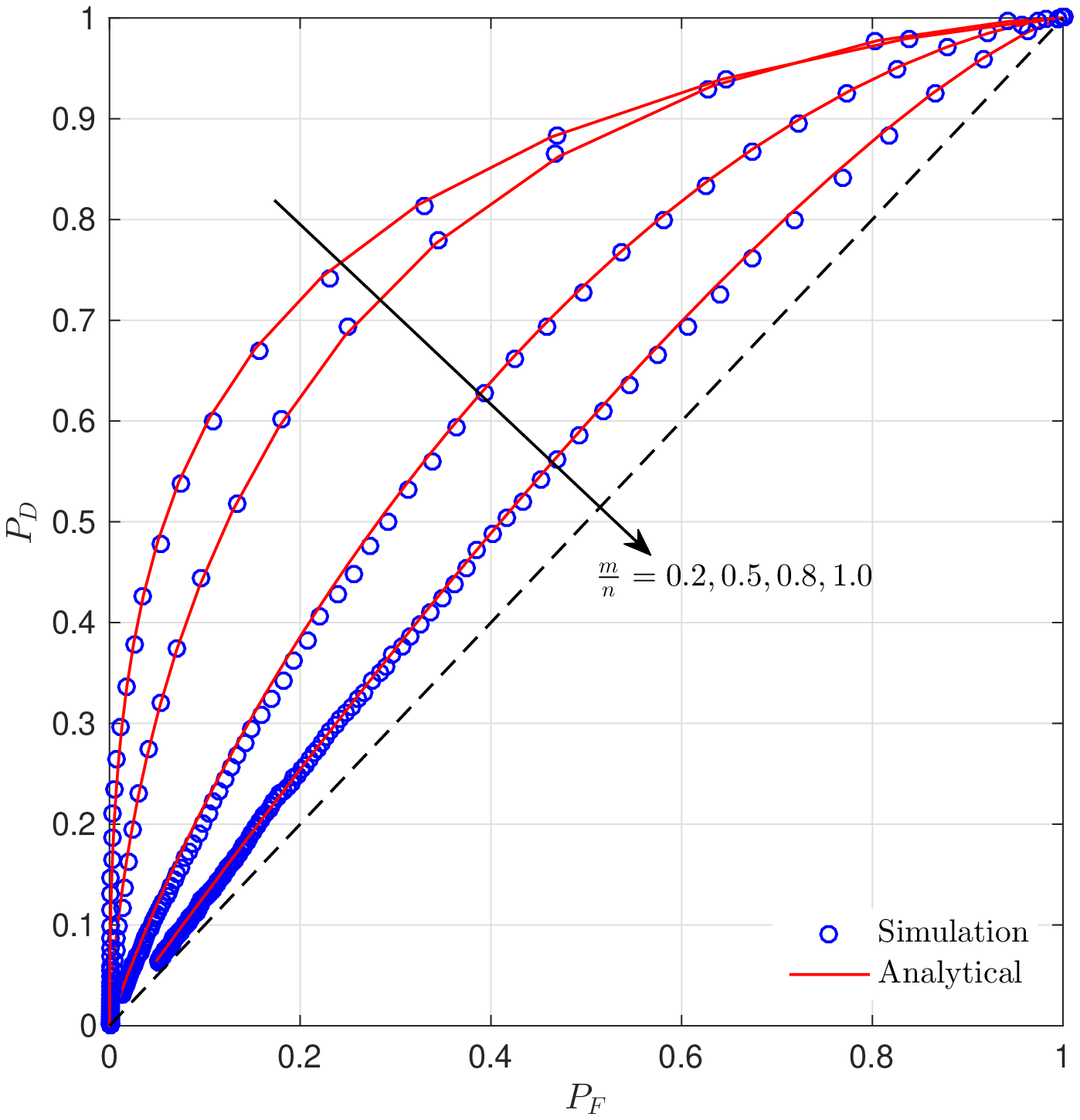}} 
	\subfloat[For different $m/p$ values with $m/n=1$ and $p=10$.]{
		\label{ROCscene4}
		\includegraphics[width=0.5\textwidth]{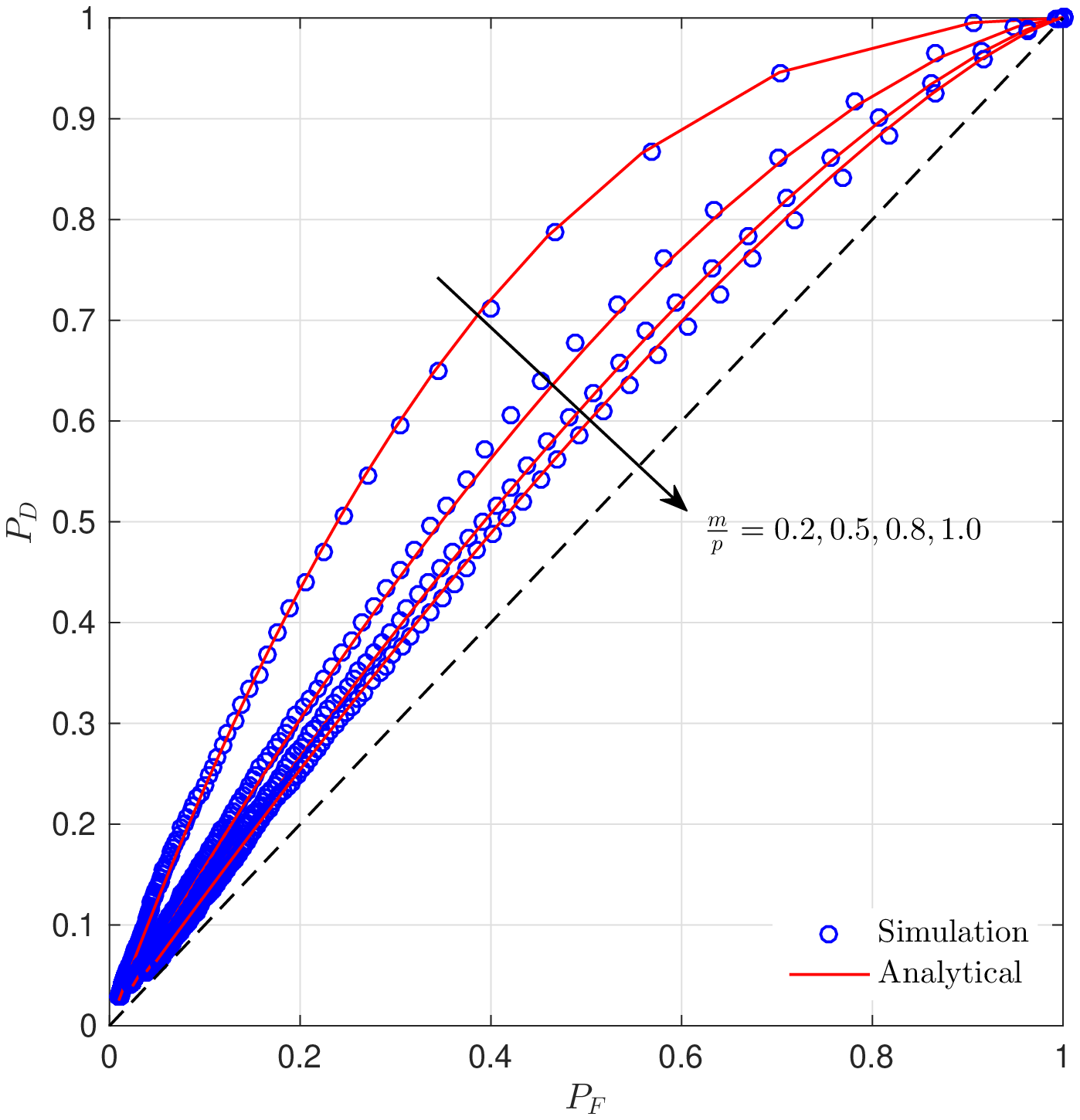}}     
	\caption{$P_D$ vs $P_F$ for different $(m,n,p)$ configurations with  $\gamma=5$\,dB.}
	\label{ROCscene34}
\end{figure}

The ROC curves corresponding to different parameter settings are shown in Figs. \ref{ROCscene12} and 
\ref{ROCscene3} and \ref{ROCscene34} depicts the power profile as a function of SNR for different $P_F$ values. As can be seen, for a fixed $P_F$, the power increase with the SNR, which is consistent with our intuition. The ROC of  maximum eigenvalue  based detection  is shown in Fig. \ref{ROCscene2} for several  SNR ($\gamma$) values, which clearly shows  that  ROC profile improves  with the increasing SNR. Since the next important parameter determining the  ROC profile is the dimensionality of the covariance matrices, we  investigate its  effect on the ROC profile. To this end, Fig. \ref{ROCscene3} shows the effect of $m/n$ for $m/p=1$. As can be seen, the disparity between $m$ and $n$ improves the ROC profile. The reason behind this observation is that the quality of the sample covariance matrix is improved when the length of the data record ($n$) increases in comparison with the dimensionality of the receiver ($m$). Since the minimum requirement for  $\bf{\widehat{R}}$ to be invertible is $m=n$, we can observe the worst ROC performance corresponds to $m/n=1$. Therefore, the effect of $m/p$ on the ROC for $m/n=1$  is shown in Fig. \ref{ROCscene4}.  As can be seen, for constant $p$, increasing $m$ degrades the ROC profile. Since we have a closed-form ROC equation for $m/n=1$, we  conduct a deeper investigation on the joint effect of $m$ and $p$ on the ROC. 

The joint effect of $m$ and $p$ is characterized in two scenarios. In particular, we consider i) varying $p$ for fixed $m$ and ii) $m$ and $p$
 both vary such that $m/p=\nu$, where $\nu>0$ is a constant. Since $p$ and $m$ take integer values only, the analysis is intractable. To circumvent this difficulty, we let $p$ and $m$ be continuous.  We  can thus write the derivative of $P_D$ with respect to $p$ as
 \begin{align*}
\frac{1}{(1-P_D)} \frac{{\rm d} P_D}{{\rm d} p}=
 \ln\left(1+\gamma-\gamma (1-P_F)^{1/mp}\right)+
 \gamma \frac{(1-P_F)^{1/mp}\ln(1-P_F)^{1/mp}}{1+\gamma-\gamma (1-P_F)^{1/mp}},
 \end{align*}
 from which we obtain using the inequality $\ln z\geq 1-1/z$, $ \frac{{\rm d} P_D}{{\rm d} p}>0$. This in turn reveals that $P_D$ increases with $p$ for all $\gamma$ and $P_F$, which is consistent with our intuition. The next immediate question of whether $P_D$ is bounded as $p\to\infty$ is answered in the sequel.
 
 We  now focus on the second scenario. As such, noting that $m/p=\nu$, we can write derivative of $P_D$ as a function of $p$ to yield
 \begin{align*}
 \frac{1}{(1-P_D)}\frac{{\rm d} P_D}{{\rm d} p}=\ln\left(1+\gamma-\gamma(1-P_F)^{1/\nu p^2}\right)+2\gamma \frac{(1-P_F)^{1/\nu p^2} \ln\left(1-P_F\right)^{1/\nu p^2}}{\left(1+\gamma-\gamma (1-P_F)^{1/\nu p^2}\right)}.
 \end{align*}
 A careful inspection of the right hand expression reveals that it has only one stationary point. However, 
 the direct evaluation of the stationary point based on the above expression does not yield any closed-form solution. Therefore, to gain insights into the $p$ value which maximizes/minimizes $P_D$, in what follows, we derive a tight bound for the stationary point. To this end, first we concentrate on the $p$ values for which $\frac{{\rm d} P_D}{{\rm d} p}<0$ for all $\gamma$ and $P_F$. As such, we use the inequalities \cite{vu17162}
 \begin{align*}
 \ln (1+z)< \frac{z(z+2)}{2(z+1)},\;\;\; z> 0,
 \end{align*}
 and $z\ln z< z(z-1)$, $z>0$ to obtain
 \begin{align*}
 & \ln\left(1+\gamma-\gamma(1-P_F)^{1/\nu p^2}\right)+2\gamma \frac{(1-P_F)^{1/\nu p^2} \ln\left(1-P_F\right)^{1/\nu p^2}}{\left(1+\gamma-\gamma (1-P_F)^{1/\nu p^2}\right)}\\
 & \qquad \qquad \qquad < 
 \frac{\gamma (1-(1-P_F)^{1/\nu p^2})}{2\left(1+\gamma-\gamma (1-P_F)^{1/\nu p^2}\right)}\left((\gamma+2)-(\gamma+4)(1-P_F)^{1/\nu p^2}\right).
 \end{align*}
 Therefore, $\frac{{\rm d} P_D}{{\rm d} p}<0$ is strict in the regime where
 \begin{align}
 \label{eqppositive}
 p>\sqrt{\frac{-\ln(1-P_F)}{-\nu\ln\left(\frac{\gamma+2}{\gamma+4}\right)}}.
 \end{align}
 
\begin{figure}[t]
		\centering
		\includegraphics[width=0.6\textwidth]{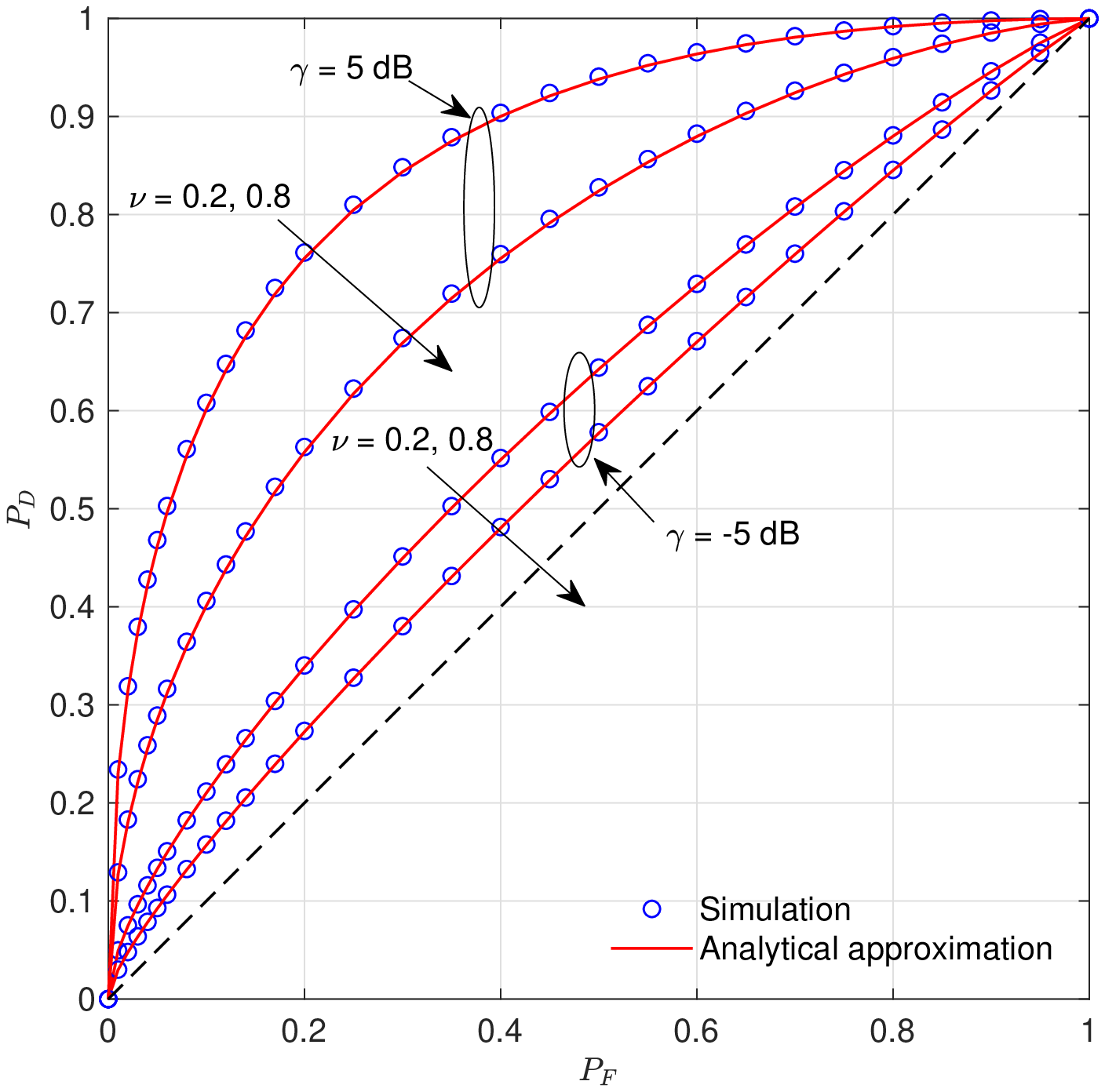}
		\caption{$P_D$ vs $P_F$ for the optimal $p$ and approximated $p$.}
		\label{p_approx}
	\end{figure}

 Again, using the inequalities \cite{vu17162}, $\ln(1+z)>2z/(2+z),\; z> 0$ and $\ln z>(1-z)/\sqrt{z},\; 0<z<1$, we have 
 \begin{align*}
  & \ln\left(1+\gamma-\gamma(1-P_F)^{1/\nu p^2}\right)+2\gamma \frac{(1-P_F)^{1/\nu p^2} \ln\left(1-P_F\right)^{1/\nu p^2}}{\left(1+\gamma-\gamma (1-P_F)^{1/\nu p^2}\right)}\\
  & >2\gamma (1-(1-P_F)^{1/\nu p^2}
  \left(\frac{1}{2+\gamma-\gamma\left(1-P_F\right)^{1/\nu p^2}}-\frac{\left(1-P_F\right)^{1/2\nu p^2}}{1+\gamma-\gamma (1-P_F)^{1/\nu p^2}}\right).
 \end{align*}
 This in turn gives that $\frac{{\rm d} P_D}{{\rm d} p}>0$ for
 \begin{align}
 \label{eqpnegative}
 p<\sqrt{\frac{-\ln(1-P_F)}{-2\nu\ln\left(\frac{\gamma+1}{\gamma+2}\right)}}.
\end{align}
Thus, we conclude that $P_D$ attains its maximum at $p=p^*$, where
\begin{align}
\sqrt{\frac{-\ln(1-P_F)}{-2\nu\ln\left(\frac{\gamma+1}{\gamma+2}\right)}}<p^*<\sqrt{\frac{-\ln(1-P_F)}{-\nu\ln\left(\frac{\gamma+2}{\gamma+4}\right)}}.
\end{align}
Having obtained the upper and lower bounds on $p^*$, a good approximation of $p^*$ can be written as\footnote{In general any convex combination of the upper and lower bounds can be a candidate for the  $p^*$.}
\begin{align}
\label{eq pmax approx}
p^*\approx \frac{1}{2}\left(\sqrt{\frac{-\ln(1-P_F)}{-\nu\ln\left(\frac{\gamma+2}{\gamma+4}\right)}}+\sqrt{\frac{-\ln(1-P_F)}{-2\nu\ln\left(\frac{\gamma+1}{\gamma+2}\right)}}\right).
\end{align}
To further highlight the accuracy of the proposed approximation, in Fig.~\ref{p_approx} we compare the optimal ROC profiles evaluated based on (\ref{eq pmax approx}) and by numerically optimizing (\ref{rocbalanced}). As can be seen from the figure, the disparity between the proposed approximation and the exact optimal solution is insignificant. Therefore, when $m=n$, under the second scenario, we can choose $p$ as per (\ref{eq pmax approx}) for fixed $P_F$, $\gamma$, and $\nu$ in view of maximizing the detection probability.



    

The detection of a very weak signal embedded in noise is particularly challenging. In this respect, it is of paramount importance to understand the behavior of $P_D$ as a function of SNR in the low SNR regime. To this end, we need to analytically characterize $P_D$ around $\gamma=0$, which is the focus of Corollary \ref{lowsnr}.
\begin{cor} \label{lowsnr} 
As $\gamma\to 0$, for a fixed value of $P_F$, $P_D(\gamma)$ admits the following form
\begin{align}
\label{eq first order}
P_D(\gamma)=\left\{\begin{array}{ll}
P_F+p  R_\epsilon(1-P_F) \gamma+o(\gamma) & \text{if $n>m$}\\
P_F+p\left[1-\left(1-P_F\right)^{1/mp}\right]\left(1-P_F\right)\gamma +o(\gamma) & \text{if $n=m$},
\end{array}\right.
\end{align}
where
\begin{align}
R_\epsilon(z)=z-\left(\frac{p+n}{p+m}\right)\frac{G(z)}{1+G(z)}z+\mathcal{K}(m,p,\alpha)
\frac{(p+n)!}{(p+m+1)!}& \left(\frac{G(z)}{1+G(z)} \right)^{m(m+\alpha+\beta+m)+1}\nonumber\\
& \qquad \times \text{det}\left[h_{i,j}\left(G(z)\right)\right]
\end{align}
with
\begin{align*}
h_{i,j}(z)=\left\{\begin{array}{ll}
\Psi_{1,j+1}(z) & i=1;\; j=1,2,\ldots,\alpha\\
\Psi_{i+1,j+1}(z) & i=2,3,\ldots,\alpha;\; j=1,2,\ldots,\alpha
\end{array}\right.
\end{align*}
and $G(z)$ being the inverse function of $F_{\lambda_\max}^{(\alpha)}(z;0)$.
\end{cor}
The proof simply follows by obtaining the Taylor expansion of the $P_D(\gamma)$ in the vicinity of $\gamma=0$.

Let us now examine  the factors affecting weak signal detection with the proposed scheme. Since the ROC curve for  the case $n>m$ is too complicated, we confine ourselves to the scenario $m=n$. Moreover, as we have already seen, this scenario may result in  the worst possible ROC and hence serves as a benchmark. Therefore, any improvement in this case  will further enhance other  ROC curves. Clearly, for very low SNR values, the most critical factor which determines the power is the coefficient of $\gamma$ given by $p\left[1-\left(1-P_F\right)^{1/mp}\right]\left(1-P_F\right)$. Since this coefficient depends on two parameters $m$ and $p$ for fixed $P_F$, we investigate the power profile when these parameters are related as follows:  i) fixed $m$, $p$ varies, ii) $m$ and $p$ both vary such that $m/p=k\in (0,1]$, and iii) $m$ and $p$ both vary such that $p-m$ is a constant. It is easy to show that under the above both options (ii) and (iii), the coefficient degrades when we increase both $p$ and $m$. In contrast, when $m$ is fixed, the coefficient gradually improves when we increase $p$. To show this, we  rewrite the above coefficient, omitting the factor $(1-P_F)$, as a function of $p$ to yield
\begin{align*}
a(p)=p\left[1-\left(1-P_F\right)^{1/mp}\right]. 
\end{align*}
Now we treat $p$ as a continuous variable and differentiate  $a(p)$ over  $p$ to yield
\begin{align*}
\frac{{\rm d}}{{\rm d}p} a(p)=\left(1-P_F\right)^{1/mp} \ln \left(1-P_F\right)^{1/mp}+1-\left(1-P_F\right)^{1/mp}.
\end{align*}
Nothing the inequality, $\ln z \geq 1-1/z$, we can easily show that $\frac{{\rm d}}{{\rm d}p} a(p)\geq 0$ for all $p,m$. This in turn establishes that $a(p)$ is a non decreasing function of $p$. The next natural question is whether there exist an upper bound for $a(p)$ as $p$ grows large. A simple limiting argument involving L'H\^opital's rule will then give
\begin{align}
\lim_{p\to\infty}a(p)=-\frac{1}{m}\ln\left(1-P_F\right).
\end{align}
Therefore, we can conclude that a power enhancement is expected in the low SNR regime if we increase $p$ for fixed $m$ and $P_F$. In particular, in the  low SNR regime (i.e., as $\gamma\to 0$), we have
\begin{align}
P_F<P_D(\gamma)< P_F-\frac{\left(1-P_F\right)}{m}\ln\left(1-P_F\right)\gamma +o(\gamma).
\end{align}

\begin{figure}
	\centering
	\subfloat[For $m=p=10$ and $n=15$.]{
		\label{generalnmsmallgamma}
		\includegraphics[width=0.5\textwidth]{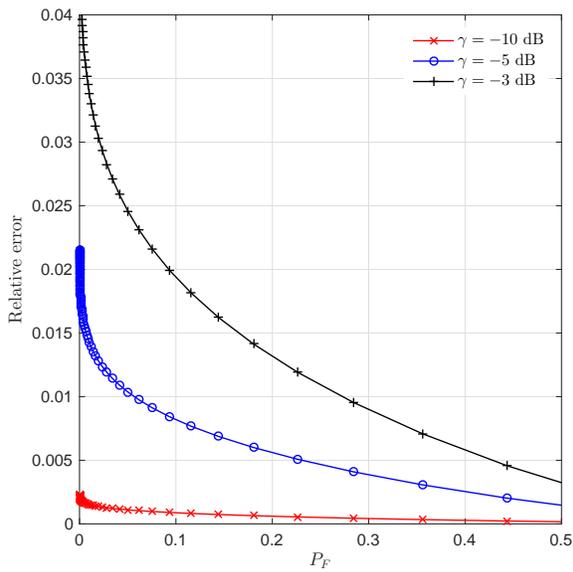}} 
	\subfloat[For $m=n=10$ and $p=20$.]{
		\label{1ROCscene5}
		\includegraphics[width=0.5\textwidth]{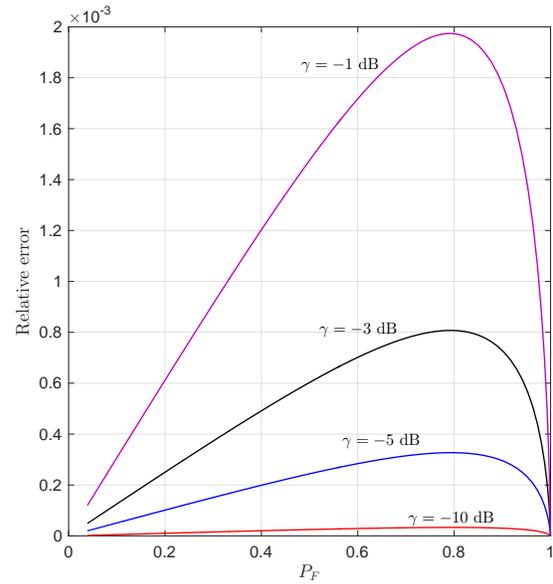}}     
	\caption{Relative error vs $P_F$ for small values of $\gamma$.}
	\label{ROCscene56}
\end{figure}

\begin{figure}
		\centerline{\includegraphics[width=0.6\figwidth]{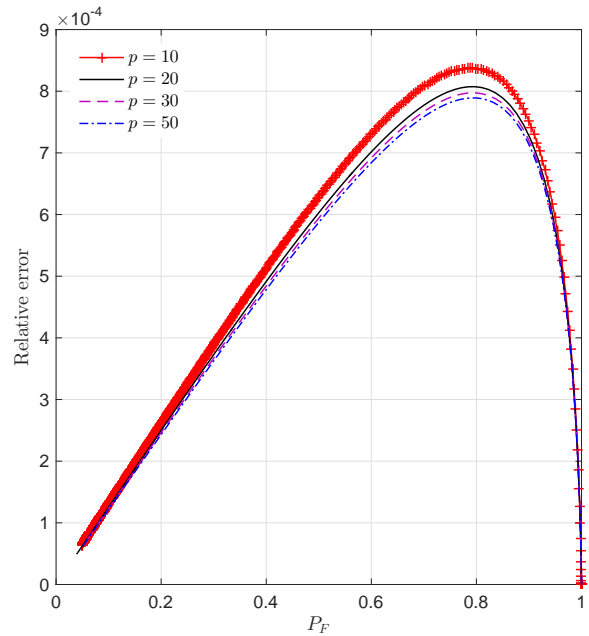}}
		\caption{Relative error vs $P_F$ for different values of $p$ with $m=n=10$ and $\gamma=-3$ dB.}
		\label{2ROCscene5}
\end{figure}

To further asses the quality of the derived first order approximations, here we numerically evaluate the relative error between the exact $P_D(\gamma)$ and the corresponding first order expansions given in (\ref{eq first order}). To be precise, we define the relative error as
\begin{align*}
\text{RE}=\frac{P_D(\gamma)-P_D^{\text{f.o.}}(\gamma)}{P_D(\gamma)}
\end{align*}
where $P_D^{\text{f.o.}}(\gamma)$ stands for the first order expansions give in (\ref{eq first order}). Figure \ref{generalnmsmallgamma} depicts the behavior of the relative error as a function of $P_F$ for a set of small values of $\gamma$. The other parameters have been chosen as $m=n=10$ and $p=15$.  Fig. \ref{generalnmsmallgamma} shows that  the diminishing $\gamma$ improves the relative error, which is anticipated. Fig. \ref{1ROCscene5} shows the relative error versus $P_F$ curve for a set of small values of $\gamma$ when $m=n=10$ and $p=20$. Although we can observe the general trend of improving relative error with the diminishing $\gamma$, for a given $\gamma$, the relative error is maximized  at a certain value of $P_F$. However, the analytical determination of this value seems an arduous task. The relative error improvement in the case of increasing $p$ is depicted in Fig. \ref{2ROCscene5}. It is interesting to observe that the relative error does not deviate much from the corresponding asymptotic limit even for finite small values of $p$ when $\gamma$ is moderately low.   

Having completed the finite-dimensional analysis, we now examine  the ROC behavior  in the asymptotic regime.


    


\subsection{Asymptotic Analysis}

Here we analyze the ROC profile in three  important asymptotic regimes. In particular, we consider the following three regimes
\begin{enumerate}
\item As $m,n,p \to \infty$ such that $\alpha, \beta$ and $\gamma$ are fixed,
\item As $p\to \infty$ such that $m=n$, and $\gamma$ are fixed,
\item As $m,n,p,\gamma \to \infty $ such that $\frac{m}{n}\to 1$, $\frac{m}{p}\to c\in(0,1]$, and $\frac{\gamma}{m}\to \theta\geq 0$.
\end{enumerate}

Following Theorem \ref{thmAsymp}, we can easily see that the maximum eigenvalue has no detection power in the first regime. Therefore, we now turn our attention to the second and third regimes. The asymptotic ROC pertaining to the second scenario can be obtained with the help of Corollary \ref{corasybalanced} as
\begin{align}
\label{eq_asyp_roc}
P_D^{\text{Asyp}}(\gamma)=\lim_{p\to\infty}P_D(\gamma)=1-
\left(1-P_F\right)^{1+\frac{\gamma}{m}}.
\end{align}
It is noteworthy that this convergence is uniform in $\gamma$. Asymptotic ROC corresponding to the third regime, is given by the following corollary
\begin{cor}
\label{corasypower}
As $m,n,p,\gamma \to \infty $ such that $\frac{m}{n}\to 1$, $\frac{m}{p}\to c\in(0,1]$, and $\frac{\gamma}{m}\to \theta\geq 0$, the ROC admits the following asymptotic limit
\begin{align}
P_D^{\text{Asy}}(\theta)=1-\left(1-P_F\right)^{1+\theta}.
\end{align}
\end{cor}
Since the above asymptotic ROC profile is independent of $c$, this expression should be valid for $c=0$ as well. Therefore, we can extend the domain of $c$ such that $c\in[0,1]$.
Clearly, when $\theta=0$ ($\gamma $ does not scale with $m$), the maximum eigenvalue has no detection power in the high dimension. This is consistent with what has been reported in \cite{Johnstone2018arx} on the power of the maximum eigenvalue below the phase transition. In contrast, when $\gamma $ scales with $m$, in the high dimension, the maximum eigenvalue still retains its detection power. For instance, when $\theta\to 0$ (the signal component is extremely weak), we have
\begin{align}
P_D^{\text{Asy}}(\theta)=P_F-(1-P_F)\ln(1-P_F)\theta +o(\theta).
\end{align}
This valuable insight is of paramount importance in detecting signals over fading channels. For instance, for Rayleigh fading, which is the most commonly used statistical model in  the literature, $\mathbf{h}$ takes the form $\mathbf{h}\sim\mathcal{CN}_m\left(\mathbf{0},\mathbf{I}_m\right)$. Now, by invoking the strong law of large numbers, we obtain
\begin{align}
\lim_{m\to\infty }\frac{||\mathbf{h}||^2}{m}\to 1, \;\; \text{almost surely}.
\end{align}
This in turn shows that $\gamma\propto m$ as $m\to \infty$ for Rayleigh fading channels. This is a clear testament to the utility of our new asymptotic ROC profile given in Corollary \ref{corasypower} in wireless applications. 

    \begin{figure}
		\centerline{\includegraphics[width=0.6\figwidth]{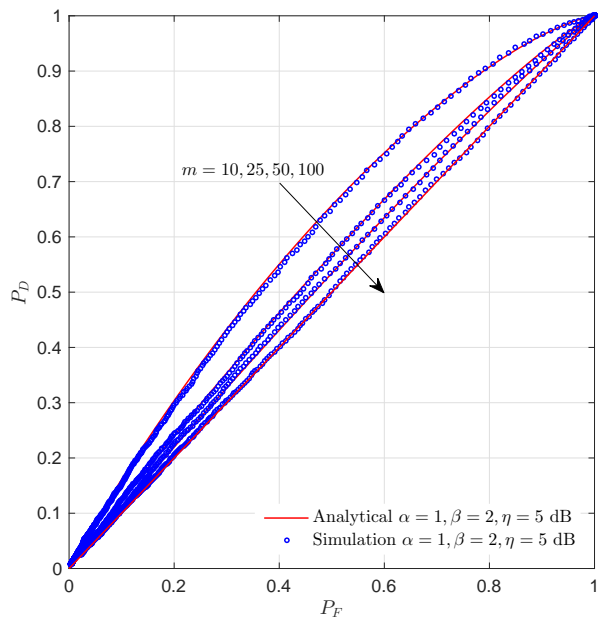}}
		\caption{$P_D$ vs $P_F$ as $m,n,p\to\infty$ such that $\alpha=n-m=1$, $\beta=p-m=2$, and $\gamma=5$ dB are fixed.}
		\label{eq_newfig}
	\end{figure}
 \begin{figure}
		\centerline{\includegraphics[width=0.6\figwidth]{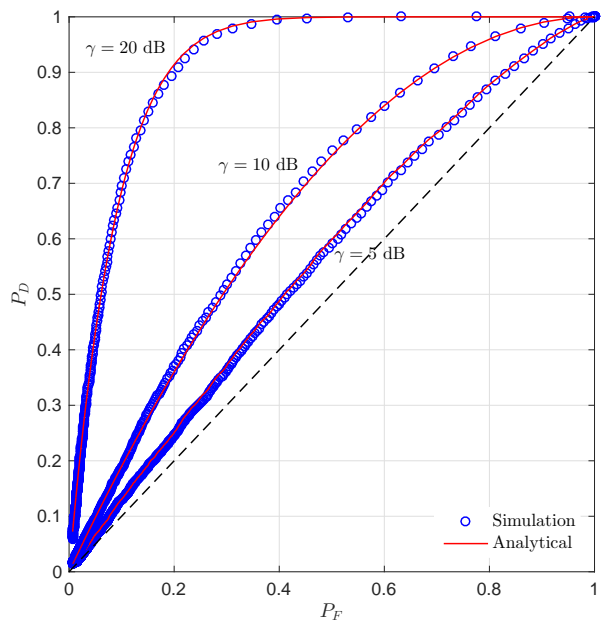}}
		\caption{Comparison of asymptotic  and finite dimensional ROC profiles corresponding to Case $2$ for different values of $\gamma$ with $m=n=10$ and $p=25$.}
		\label{eq31_2}
	\end{figure}

\begin{figure}
		\centerline{\includegraphics[width=0.6\figwidth]{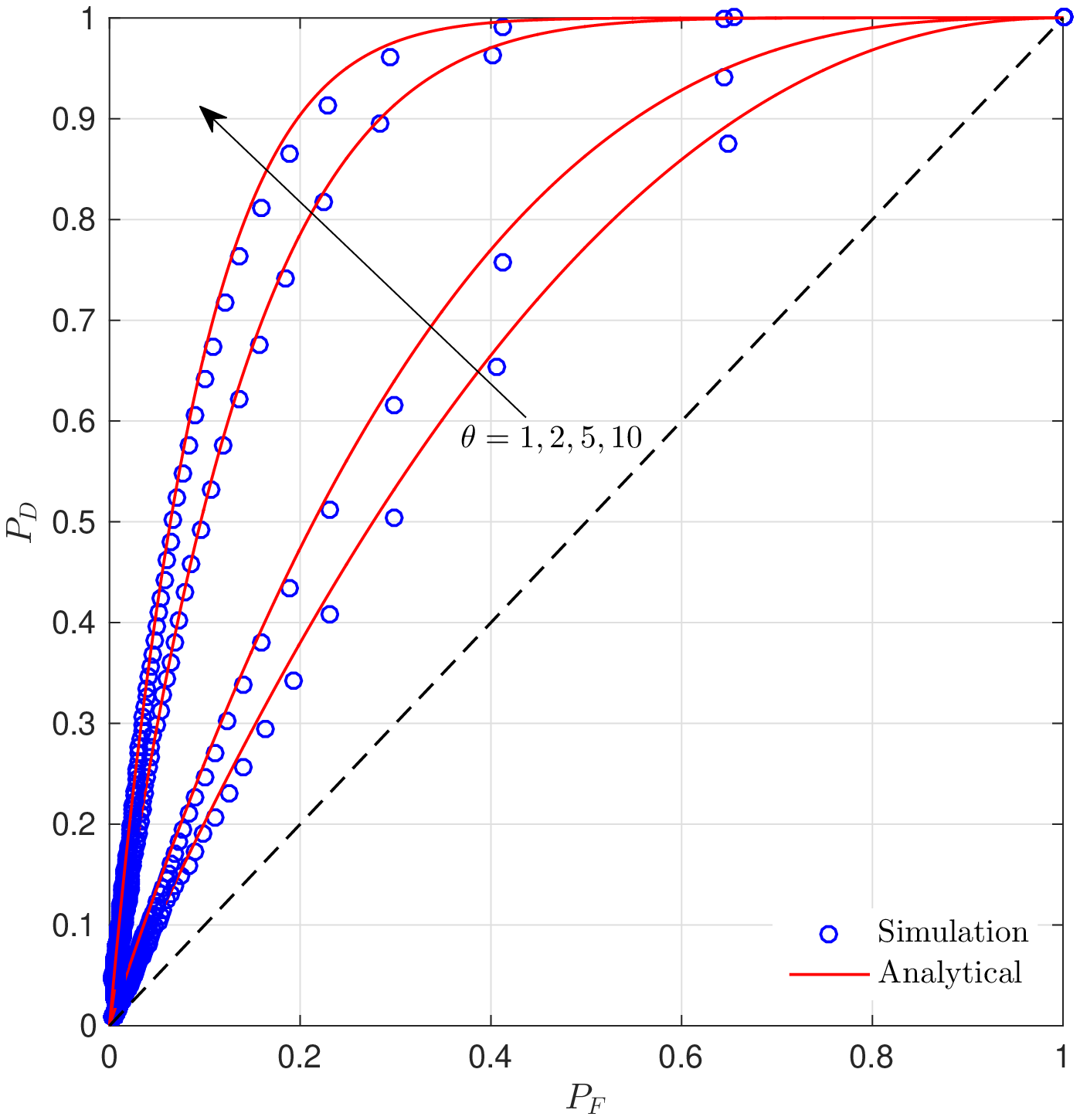}}
		\caption{Comparison of asymptotic  and finite dimensional ROC profiles corresponding to Case $3$ for different values of $\theta$ with $m=n=25$ and $c=1$.}
		\label{eq32_2}
	\end{figure}
	

The above dynamics are depicted in Figs. \ref{eq_newfig}, \ref{eq31_2}, and \ref{eq31_2}. In particular, Fig. \ref{eq_newfig} compares the analytical ROC profiles with the numerical results for an increasing sequence of $m$ values when $\alpha=1, \beta=2$, and $\gamma=5\; {\rm dB}$ are fixed. As can be seen from the figure, when $m$ increases the ROC profiles go arbitrary closer to $P_D=P_F$ curve, thereby demonstrating the loss of the power of the test. This observation is consistent with what we have analytically shown related to the regime where $\alpha$ and $\beta$ are fixed with $\gamma=5\; {\rm dB}$. The effect of increasing $p$ on the ROC profile is depicted in Fig. \ref{eq31_2}. The analytical curves are based (\ref{eq_asyp_roc}) and the close matching between the analytical and simulation results can be seen from the figure. This in turn shows us that that the analytical asymptotic result (as $p\to\infty$) derived in (\ref{eq_asyp_roc})  serves as a good approximation to finite values of $p$ as well. Finally, Fig. \ref{eq31_2} compares the analytical asymptotic result for the third region where $m,n,p,\gamma \to \infty $ such that $\frac{m}{n}\to 1$, $\frac{m}{p}\to c\in(0,1]$, and $\frac{\gamma}{m}\to \theta\geq 0$ with the simulation results. Again, closely matching two results reveal that our asymptotic analytical expression serves as a good approximation to the finite dimensional case as well. These results clearly indicate that, when $\gamma$ scales with $m$, the maximum eigenvalue retains its detection power, whereas it looses the detection power when $\gamma$ does not scale with $m$.     
\section{Conclusion}
This paper investigates the signal detection problem in colored noise with unknown covariance matrix. Thus, the presence of a signal is detected by using the maximum generalized eigenvalue of the  whitened sample covariance matrix. Equivalently, we need to determine the distribution of  the maximum eigenvalue of the deformed Jacoby unitary ensemble. To this end, we exploited the  powerful orthogonal polynomial approach to  develop a new  c.d.f. expression  of the maximum  eigenvalue of the deformed JUE. Subsequently, we used it  to determine the ROC of the detector. It turns out that, for a fixed SNR, when $m$ (i.e., the dimensionality of the detector), $n$ (i.e., the number of noise-only samples), and $p$ (i.e., the number of signal-plus-noise samples) increase over finite values such that $m=n$ and $m/p$ is constant, we obtain an optimal ROC profile corresponding to specific $m,n$, and $p$ values. In contrast, in the above setting, when $m,p$, and $n$ increase asymptotically, the maximum eigenvalue gradually loses its detection power. This is not surprising, since under the above asymptotic setting, the detector operates below the so called phase transition where the maximum eigenvalue has no detection power. However, when the SNR scales with $m$, in the same asymptotic regime, the maximum eigenvalue retains its detection power. This fact is of paramount importance in detecting a signal in colored noise over fading channels (Rayleigh fading) where the SNR scales with the dimensionality of the system. Clearly, $m=n$ is the minimum requirement for the noise-only covariance matrix to be full rank (or nearly rank deficient). Therefore, some of the key results developed in this paper related to the setting $m=n$ shed some light into the regime where noise-only covariance matrix is nearly rank deficient. However, the analysis pertaining to the regime where the latter matrix is fully rank deficient remains an important open problem.  


	\appendices
	\section{Proof of the joint density of the eigenvalues}\label{appPrathapa}
	\noindent Following James \cite{meJames}, we can write the joint density of the eigenvalues of $\mathbf{W}_1\mathbf{W}_2^{-1}$ as 
	\begin{align}
	\label{pdf_decom}
	f(\lambda_1,\lambda_2,\cdots,\lambda_m)=\frac{\mathcal{K}_1(m,n,p)}{(1+\eta)^p}
	\prod_{j=1}^m \lambda_j^{p-m}\Delta_m^2(\boldsymbol{\lambda})
	\int_{U(m)}
	\frac{1}{\text{det}^\alpha[\mathbf{I}_m+\boldsymbol{\Sigma}_1^{-1}\mathbf{U}\boldsymbol{\Lambda}\mathbf{U}^\dagger]}\;  {\rm d}\mathbf{U}.
	\end{align}
	where $\alpha=p+n$ and ${\rm{d}}\mathbf{U}$ is the invariant measure on the unitary group $U(m)$, normalized to make the total measure unity. Let us now focus on simplifying the above matrix integral. To this end, we use (\ref{spike_cov}) to rewrite
	\begin{align}
	\int_{U(m)}
	\frac{1}{\text{det}^\alpha[\mathbf{I}_m+\boldsymbol{\Sigma}_1^{-1}\mathbf{U}\boldsymbol{\Lambda}\mathbf{U}^\dagger]}\;  {\rm d}\mathbf{U}&=
	\int_{U(m)}
	\frac{1}{\text{det}^\alpha[\mathbf{I}_m+\mathbf{U}\boldsymbol{\Lambda}\mathbf{U}^\dagger-\mathbf{V}\boldsymbol{\Lambda}_\eta \mathbf{V}^\dagger \mathbf{U}\boldsymbol{\Lambda}\mathbf{U}^\dagger]}\;  {\rm d}\mathbf{U}\nonumber\\
	&= \int_{U(m)}
	\frac{1}{\text{det}^\alpha[\mathbf{I}_m+\boldsymbol{\Lambda}-\mathbf{U}^\dagger\mathbf{V}\boldsymbol{\Lambda}_\eta \mathbf{V}^\dagger \mathbf{U}\boldsymbol{\Lambda}]}\;  {\rm d}\mathbf{U}
	\end{align}
	where $\boldsymbol{\bar{\Lambda}}=\boldsymbol{\Lambda}(\mathbf{I}_m+\boldsymbol{\Lambda})^{-1}=\text{diag}\left(\bar{\lambda}_m,\cdots,\bar{\lambda}_1\right)=\text{diag}\left(\frac{\lambda_m}{1+\lambda_m},\cdots,\frac{\lambda_1}{1+\lambda_1}\right)$. Therefore, after some algebra, we obtain
    \begin{align*}
    \int_{U(m)}
	\frac{1}{\text{det}^\alpha[\mathbf{I}_m+\boldsymbol{\Sigma}_1^{-1}\mathbf{U}\boldsymbol{\Lambda}\mathbf{U}^\dagger]}\;  {\rm d}\mathbf{U}&=\frac{1}{\text{det}^\alpha[\mathbf{I}_m+\boldsymbol{\Lambda}]}\int_{U(m)}
	\frac{1}{\text{det}^\alpha[\mathbf{I}_m-\mathbf{H}\boldsymbol{\Lambda}_\eta\mathbf{H}^\dagger\boldsymbol{\bar{\Lambda}}]}{\rm d}\mathbf{H} 
    \end{align*}
 where ${\rm d}\mathbf{H}$ is the invariant measure on the unitary group $U(m)$, normalized to make the total measure unity.
    Since $\boldsymbol{\Lambda}_\eta$ is rank one, we can further simplify the above matrix integral to yield
	\begin{align}
	\int_{U(m)}
	\frac{1}{\text{det}^\alpha[\mathbf{I}_m+\boldsymbol{\Sigma}_1^{-1}\mathbf{U}\boldsymbol{\Lambda}\mathbf{U}^\dagger]}\;  {\rm d}\mathbf{U}&=
	\frac{1}{\text{det}^\alpha[\mathbf{I}_m+\boldsymbol{\Lambda}]}\int_{U(m)}
	\frac{1}{\left(1-\text{tr}\left(\mathbf{H}\boldsymbol{\Lambda}_\eta\mathbf{H}^\dagger\boldsymbol{\bar{\Lambda}}\right)\right)^\alpha}\;  {\rm d}\mathbf{H}.
	\end{align}
	Now it is worth observing that
	\begin{align}
	\text{tr}\left(\mathbf{H}\boldsymbol{\Lambda}_\eta\mathbf{H}^\dagger\boldsymbol{\bar{\Lambda}}\right)=\frac{\eta}{1+\eta}\mathbf{h}_1\boldsymbol{\bar{\Lambda}}\mathbf{h}_1^\dagger\leq \frac{\eta}{1+\eta} \frac{\lambda_m}{1+\lambda_m}<1.
	\end{align}
	This in turn enables us to utilize the relation
	\begin{align}
	\frac{1}{s^\alpha}=\frac{1}{\Gamma(\alpha)}\int_0^\infty y^{\alpha-1} e^{-sy }{\rm d}y,\;\;\; s>0
	\end{align}
	to express the above matrix integral as
	\begin{align}
	\label{int_split}
	\int_{U(m)}
	\frac{1}{\text{det}^\alpha[\mathbf{I}_m+\boldsymbol{\Sigma}_1^{-1}\mathbf{U}\boldsymbol{\Lambda}\mathbf{U}^\dagger]}\;  {\rm d}\mathbf{U}=
	\frac{1}{\text{det}^\alpha[\mathbf{I}_m+\boldsymbol{\Lambda}]}\frac{1}{\Gamma(\alpha)}
	\int_0^\infty y^{\alpha-1}e^{-y} \Phi(y) {\rm d}y
	\end{align}
	where
	\begin{align}
	\Phi(y)=
	\int_{U(m)}
	e^{y\text{tr}\left(\mathbf{H}\boldsymbol{\Lambda}_\eta\mathbf{H}^\dagger\boldsymbol{\bar{\Lambda}}\right)} {\rm d}\mathbf{H} 
	\end{align}
	and we have taken the liberty of changing the order of integration. Noting the fact that
	\begin{align}
	e^{y\text{tr}\left(\mathbf{H}\boldsymbol{\Lambda}_\eta\mathbf{H}^\dagger\boldsymbol{\bar{\Lambda}}\right)}=
	{}_0\widetilde{F}_0(y\mathbf{H}\boldsymbol{\Lambda}_\eta\mathbf{H}^\dagger\boldsymbol{\bar{\Lambda}})
	\end{align}
	we may use the splitting formula [eq. 92, James] to yield
	\begin{align}
	\Phi(y)=\int_{U(m)} {}_0\widetilde{F}_0(y\mathbf{H}\boldsymbol{\Lambda}_\eta\mathbf{H}^\dagger\boldsymbol{\bar{\Lambda}}) {\rm d}\mathbf{H}
	={}_0\widetilde{F}_0\left(y\boldsymbol{\Lambda}_\eta,\boldsymbol{\bar{\Lambda}}\right).
	\end{align}
	Following \cite{meWang}, we can show that
	\begin{align}
	{}_0\widetilde{F}_0\left(y\boldsymbol{\Lambda}_\eta,\boldsymbol{\bar{\Lambda}}\right)=\Gamma(m)\left(\frac{1+\eta}{\eta}\right)^{m-1}
	y^{1-m} \sum_{k=1}^m
	\frac{e^{\frac{\eta \bar{\lambda}_k}{(1+\eta)}y}}{\prod_{\substack{j=1\\
				j\neq k}} (\bar{\lambda}_k-\bar{\lambda}_j)}
	\end{align}
	from which we obtain upon substituting into (\ref{int_split}) with some algebra
	\begin{align}
	\label{matint_sol}
	\int_{U(m)}
	\frac{1}{\text{det}^\alpha[\mathbf{I}_m+\boldsymbol{\Sigma}_1^{-1}\mathbf{U}\boldsymbol{\Lambda}\mathbf{U}^\dagger]}\;  {\rm d}\mathbf{U}& =
	\frac{\Gamma(\alpha-m+1)\Gamma(m)}{\Gamma(\alpha)} \left(\frac{1+\eta}{\eta}\right)^{m-1}
	\frac{1}{\prod_{j=1}^m(1+\lambda_j)^\alpha}\nonumber\\
	& \times 
	\sum_{k=1}^m
	\frac{1}{\prod_{\substack{j=1\\
				j\neq k}} (\bar{\lambda}_k-\bar{\lambda}_j)}\frac{1}{\left(1-\frac{\eta \bar{\lambda}_k}{1+\eta}\right)^{\alpha-m+1}}.
	\end{align}
	Finally, using (\ref{matint_sol}) in (\ref{pdf_decom}) with some algebraic manipulation we obtain (\ref{jpdf}), which concludes the proof.

	\section{Proof of the c.d.f. of the maximum eigenvalue }\label{appa}
	\noindent By exploiting the symmetry, the ordered region of integration in \eqref{cdfdef} can be rearranged as an unordered region to yield
	\begin{align}
	\Pr(x_{\max}\leq t)=\frac{\mathcal{K}(m,n,p)}{(m)!\eta^{m-1}(1+\eta)^{p+1-m}} \sum_{k=1}^m\int_{\left[0,t\right]^{m}}
	\Delta_m^2(\mathbf{x})
	\frac{\prod_{j=1}^{m}x_{j}^{p-m}(1-x_{j})^{n-m}}{\prod_{\substack{j=1\\j\neq k}}^{m}(x_{i}-x_{j})\left(1-\frac{\eta}{1+\eta}x_{k}\right)^{p+n+1-m}} \text{ d}\textbf{x}
	\end{align}
    where $[0,t]^m=[0,t]\times [0,t]\times \ldots\times [0,t]$ with $\times $ denoting the Cartesian product.
    Since each term in the above summation contributes the same amount to the final solution, it can be further simplified as
	\begin{equation*}
	\Pr(x_{\max} \leq\ t)=\dfrac{\mathcal{K}}{(m-1)!} \int_{ [0,t]^{m}}^{} \Delta^{2}_{m}(\textbf{x}) \frac{\prod_{j=1}^{m}x_{j}^{\beta}(1-x_{j})^{\alpha}}{\prod_{j=2}^{m}(x_{1}-x_{j})\left(1-\frac{\eta}{1+\eta}x_{1}\right)^{\gamma}}  \text{ d}\textbf{x}.
	\end{equation*}
	where,
	\begin{equation*}
	\mathcal{K}= \frac{\mathcal{K}(m,n,p)}{\eta^{m-1}(1+\eta)^{p+1-m}}.
	\end{equation*}
	\noindent Here we have relabeled the variables as $ \alpha = n-m$, $\beta=p-m$ and $ \gamma=m+\alpha+\beta+1 $ for notational concision. To facilitate further analysis, let us decompose the Vandermonde determinant as
    \begin{align*}
    \Delta_{m}(\textbf{x}) = \prod_{j=2}^{m}(x_{1}-x_{j}) \Delta_{m-1}(\textbf{x})
    \end{align*}
	and relabel  the variables $ x_{1}=y$ and $ x_{j} = z_{j-1}$, $j=2, 3, ..., m$, to obtain
	\begin{equation}\label{eq2}
	\Pr(x_{\max} \leq t)=\dfrac{\mathcal{K}}{(m-1)!} \int_{[0,t]^{m}}^{}\frac{y^{\beta}(1-y)^{\alpha}}{\left(1-\frac{\eta}{1+\eta}y\right)^{\gamma}}\prod_{j=1}^{m-1}z_{j}^{\beta}(1-z_{j})^{\alpha}(y-z_{j}) \Delta_{m-1}^{2}(\textbf{z}) \text{ d}\textbf{z}
	\end{equation}
where $ \textbf{z} \in \mathbb{R}^{m-1} $. Now we apply the variable transformations $ y = tx $ and $ z_{j} = ts_{j} $, $ j=1,2,...,m-1 $, to make the region of integration independent of $t$ in (\ref{eq2}). Consequently we have after some  algebraic manipulations
	\begin{equation}\label{eq3}
	\Pr(x_{\max} \leq t)=\dfrac{\mathcal{K}}{(m-1)!} t^{m(\beta+m-1)+1} \int_{0}^{1}\frac{x^{\beta}(1-tx)^{\alpha}}{\left(1-\frac{\eta t }{1+\eta}x\right)^{\gamma}} \mathcal{Q}_{m-1}(\beta,\alpha,x,t) {\rm d}x
	\end{equation}
	where,
	\begin{equation}\label{qfunc}
	\mathcal{Q}_{m}(\beta,\alpha,x,t)= \int_{[0,1]^{m}}^{} \prod_{j=1}^{m}s_{j}^{\beta}(1-ts_{j})^{\alpha}(x-s_{j}) \Delta_{m}^{2}(\textbf{s}) {\rm d}\textbf{s}.
	\end{equation}
	Following Appendix \ref{appa.1}, we can solve the above multidimensional integral to yield
	\begin{dmath}
    \label{khatriform}
		\mathcal{Q}_{m}(\beta,\alpha,x,t) = \tilde{\mathcal{C}}_{(0,\beta,m)}\frac{t^{\alpha m+1}}{2^{m(\alpha+\beta+m+1)+\frac{\alpha}{2}(\alpha+1)}\prod_{j=1}^{\alpha-1}(j)!}\frac{1}{\left(1-xt\right)^{\alpha}} \times\det\left[P_{m+i-1}^{(0,\beta)}\left(h_{\frac{1}{x}}\right)\hspace{6mm} (m+i+\beta)_{j-2}P_{m+i-j+1}^{(j-2,\beta+j-2)}\left(h_{t}\right)\right]_{\substack{i=1,2,...,\alpha+1\\j=2,3,...,\alpha+1}}
	\end{dmath}
    where
	\begin{equation}
	\tilde{\mathcal{C}}_{(0,\beta,m)} = \mathcal{C}_{(0,\beta,m)}\prod_{j=1}^{\alpha+1}2^{m+j-1}\frac{(m+j-1)!(m+\beta+j-1)!}{(2m+2j+\beta-2)!},
	\end{equation}
	\begin{dmath}
		\mathcal{C}_{(0,\beta,m)} =  2^{m(\beta+m)}\prod_{j=0}^{m-1}\frac{(j)!(j+1)!(\beta+j)!}{(\beta+m+j)!},
	\end{dmath}
    and $ h_{t} = \frac{2}{t} -1$.
    Using (\ref{khatriform}) in (\ref{eq3}) with some algebraic manipulation we have 
		\begin{align}\label{eqr6}
			\Pr(x_{\max} \leq t)&=\dfrac{\mathcal{K}\tilde{\mathcal{C}}_{(0,\beta,m-1)}t^{m(\alpha+\beta+m-1)+1}}{(m-1)!2^{(m-1)(\alpha+\beta+m)+\frac{\alpha}{2}(\alpha+1)}\prod_{j=1}^{\alpha-1}(j)!}\nonumber\\
            & \times \int_{0}^{1}\frac{x^{\beta}}{\left(1-\frac{\eta t }{1+\eta}x\right)^{\gamma}}\det\left[P_{m+i-2}^{(0,\beta)}(2x-1)  \hspace{6mm}  \Psi_{i,j}\left(\frac{t}{1-t}\right)\right]_{\substack{i=1,2,...,\alpha+1\\j=2,3,...,\alpha+1}}\text{ d}x.
		\end{align}
         Having observed that only the first column of the determinant in the integrand depends on $x$, we can rewrite the above integral as
        \begin{align}
        \label{cdf_interim}
			\Pr(x_{\max} \leq t)&=\dfrac{\mathcal{K}\tilde{\mathcal{C}}_{(0,\beta,m-1)}t^{m(\alpha+\beta+m-1)+1}}{(m-1)!2^{(m-1)(\alpha+\beta+m)+\frac{\alpha}{2}(\alpha+1)}\prod_{j=1}^{\alpha-1}(j)!}\nonumber\\
            &  \times \det\left[\int_{0}^{1}\frac{x^{\beta}}{\left(1-\frac{\eta t }{1+\eta}x\right)^{\gamma}}P_{m+i-2}^{(0,\beta)}(2x-1)  \hspace{6mm}  \Psi_{i,j}\left(\frac{t}{1-t}\right)\right]_{\substack{i=1,2,...,\alpha+1\\j=2,3,...,\alpha+1}}\text{ d}x.
		\end{align}
        For clarity, let us focus on the integral in the above equation.
    In this respect, we may use the relation \eqref{jacobidef2} followed by the variable transformation $ y=1-x $ to arrive at
	\begin{align*}
	&\int_{0}^{1}\frac{x^{\beta}}{\left(1-\frac{\eta t }{1+\eta}x\right)^{\gamma}}P_{m+i-1}^{(0,\beta)}(2x-1)\text{ d}x\\
    &\qquad \qquad \qquad \quad=	\frac{1}{\left(1-\frac{\eta t}{1+\eta}\right)^{\gamma}} \int_{0}^{1}\frac{\left(1-y\right)^{\beta}}{\left(1+\frac{\eta t}{1+\eta (1-t)}y\right)^{\gamma}}\Hypergeometric{2}{1}{-m-i+2,m+\beta+i-1}{1}{y}\text{ d}y,
    \end{align*}
   which can be solved using \cite[eq. 399.6]{Bateman1954book} to obtain
    \begin{align}
   &\int_{0}^{1}\frac{x^{\beta}}{\left(1-\frac{\eta t }{1+\eta}x\right)^{\gamma}}P_{m+i-1}^{(0,\beta)}(2x-1)\text{ d}x \nonumber\\
   &\qquad \quad =\frac{\Gamma(\beta+1)}{\Gamma(\beta+m+i)\Gamma(3-m-i)}
   \Hypergeometric{3}{2}{\beta+1,\gamma,1}{\beta+m+i,3-m-i}{\frac{\eta t }{1+\eta}}.
   \end{align}
   To facilitate further analysis, nothing that $ \frac{\eta t }{1+\eta} < 1$, we may replace the hypergeometric function with its equivalent infinite series expansion to yield
   \begin{align} &\int_{0}^{1}\frac{x^{\beta}}{\left(1-\frac{\eta t }{1+\eta}x\right)^{\gamma}}P_{m+i-1}^{(0,\beta)}(2x-1)\text{ d}x \nonumber\\
		&\qquad \quad=\frac{\Gamma(\beta+1)}{\Gamma(\beta+m+i)\Gamma(3-m-i)}
		\sum_{k=0}^{\infty}\frac{(\beta+1)_{k}(\gamma)_{k}(1)_{k}}{k!(\beta+m+i)_{k}(3-m-i)_{k}}\left(\frac{\eta t }{1+\eta}\right)^{k}.
	\end{align}
	Since the Gamma function has poles at negative integer values including zero, the above series is nonzero if the argument of $\Gamma\left(3-m-i+k\right)=\Gamma(3-m-i)\left(3-m-i\right)_k$ is a positive integer. To this end, $k$ should satisfy the inequality $k\geq m+i-2$. Therefore, by relabeling summation index $k$ as $j=k-m-i+2$, we obtain 
	\begin{dmath}
		\int_{0}^{1}\frac{x^{\beta}}{\left(1-\frac{\eta t}{1+\eta}x\right)^{\gamma}}P_{m+i-2}^{(0,\beta)}(2x-1)\text{ d}x = \frac{\Gamma(\beta+1)}{\Gamma(\beta+m+i)}
		\sum_{j=0}^{\infty}\frac{(\beta+1)_{m+i+j-2}(\gamma)_{m+i+j-2}(1)_{m+i+j-2}}{(m+i+j-2)!(\beta+m+i)_{m+i+j-2}\Gamma(j+1)}\left(\frac{\eta t }{1+\eta}\right)^{m+i+j-2}.
	\end{dmath}
    The above infinite series can be rearranged by using the addition formula $ (a)_{n+k}=(a)_{n}(a+n)_{k} $ with some algebraic manipulations to yield 
	\begin{align}
    \label{jacobi_integral}
		\int_{0}^{1}\frac{x^{\beta}}{\left(1-\frac{\eta t }{1+\eta}x\right)^{\gamma}}P_{m+i-2}^{(0,\beta)}(2x-1)\text{ d}x 
        =\frac{\Gamma(a_i)\Gamma(b_i)}{\Gamma(\gamma)\Gamma(c_i)}\left(\frac{\eta t}{1+\eta}\right)^{m+i-2}
   \Hypergeometric{2}{1}{a_i,b_i}{c_i}{\frac{\eta t}{1+\eta}},
	\end{align}
    where $a_i=\beta+m+i-1$, $b_i=\gamma+m+i-2$, and $c_i=\beta+2m+2i-2$.
Now we substitute (\ref{jacobi_integral}) into (\ref{cdf_interim}) followed by some algebraic manipulations to obtain the c.d.f. of $x_\max$ as
	\begin{align}\label{cdfXmaxFinite}
		\Pr(x_{\max} \leq t)
		&=\dfrac{t^{m(\alpha+\beta+m)}}{(p-1)!(1+\eta)^{p}}\left(\prod_{j=0}^{\alpha-1}\dfrac{(p+m+j-1)!}{(p+m+2j)!}\right)\nonumber\\
		& \times \det\left[\frac{\Gamma(a_i)\Gamma(b_i)}{\Gamma(c_i)}\left(\frac{\eta t}{1+\eta}\right)^{i-1}\Hypergeometric{2}{1}{a_i,b_i}{c_i}{\frac{\eta t}{1+\eta}}\hspace{6mm} \Psi_{i,j}\left(\frac{t}{1-t}\right)\right]_{\substack{i=1,2,...,\alpha+1\\j=2,3,...,\alpha+1}}.
	\end{align}
Now (\ref{cdfthm}) with $\Phi(t,\eta)$ given by (\ref{eqHyper}) follows by transforming the variable $x_\max$ to $\lambda_{\max}$ using the functional relation $\lambda_\max=x_\max/(1-x_\max)$.  Finally, noting that $c_i-b_i$ is a negative integer, we may use the hypergeometric transformation \cite[eq. 15.3.4]{abramowitz1964handbook},
	\begin{dmath}
    \label{hypertransform}
		{}_2F_1(a,b,c,z)=\left(1-z\right)^{-a}{}_2F_1\left(a,c-b,c,\frac{z}{z-1}\right),
	\end{dmath}
to arrive at the finite series form of $\Phi(t,\eta)$, thereby concluding the proof.
	\section{}
	\label{appa.1}
    
    Let us change the region of integration in \eqref{qfunc} from $[0,1]^m$ to $[-1,1]^m$ by using the variable transformation  $ s_{j}=\frac{1+z_{j}}{2} $, $ j=1,2,...,m $, to yield 
	\begin{equation}\label{eq4}
	\mathcal{R}_{m}(\beta,\alpha,x,t) =  \frac{t^{\alpha m}}{2^{m(m+\beta+\alpha+1)}}\mathcal{R}_{m}(\beta,\alpha,x,t) 
	\end{equation}
	where 
	\begin{equation}
    \label{mehtaourint}
	\mathcal{R}_{m}(\beta,\alpha,x,t) = \int_{[-1,1]^{m}}^{} \prod_{j=1}^{m}(1+z_{j})^{\beta}(h_{t}-z_{j})^{\alpha}\left(h_{\frac{1}{x}}-z_{j}\right) \Delta_{m}^{2}(\textbf{z}) \text{ d}\textbf{z},
	\end{equation}
    with $ h_{t} = \frac{2}{t} -1$ and $ \textbf{z} \in \mathbb{R}^{m} $.
    Our strategy is to start with a related integral given in \cite[Eqs. 22.4.2, 22.4.11]{me11} as
    \begin{dmath} \label{eqmehta}
		\int_{[-1,1]^{m}}^{} \prod_{j=1}^{m}(1+z_{j})^{\beta}\prod_{i=1}^{\alpha+1}(r_{i}-z_{j}) \Delta_{m}^{2}(\textbf{z}) \text{ d}\textbf{z}                            
		\hspace{5mm} =  \mathcal{C}_{(0,\beta,m)}\Delta_{\alpha+1}^{-1}(\textbf{r})\det\left[C_{m+i-1}(r_{j})\right]_{i,j=1,2,...,\alpha+1}
	\end{dmath}
    where
    \begin{align*}
    \mathcal{C}_{(0,\beta,m)}=2^{m(\beta+m)}\prod_{j=0}^{m-1}\frac{(j)!(j+1)!(\beta+j)!}{(\beta+m+j)!}
    \end{align*}
    and $ C_{k}(x) $ are monic polynomials orthogonal with respect to the weight $ (1+x)^{\beta} $, over $ -1\leq x \leq 1 $. Since Jacobi polynomials are orthogonal with respect to the preceding weight, we use $ C_{k}(x) = 2^{k}\frac{(k+\beta)!(k)!}{(2k+\beta)!}P_{k}^{(0,\beta)}(x)$ in (\ref{eqmehta}) to obtain
    \begin{equation}\label{eqmehta1}
	\int_{[-1,1]^{m}}^{} \prod_{j=1}^{m}(1+z_{j})^{\beta}\prod_{i=1}^{\alpha+1}(r_{i}-z_{j}) \Delta_{m}^{2}(\textbf{z}) \text{ d}\textbf{z} = \frac{\tilde{\mathcal{C}}_{(0,\beta,m)}}{\Delta_{\alpha+1}(\textbf{r})}\det\left[P_{m+i-1}^{(0,\beta)}(r_{j})\right]_{i,j=1,2,...,\alpha+1}
	\end{equation}
	where,
	\begin{equation}
	\tilde{\mathcal{C}}_{(0,\beta,m)} = \mathcal{C}_{(0,\beta,m)}\prod_{j=1}^{\alpha+1}2^{m+j-1}\frac{(m+j-1)!(m+\beta+j-1)!}{(2m+2j+\beta-2)!}.
	\end{equation}
  In the above, $r_i$s are generally distinct parameters. Nevertheless, if we choose $r_i$ such that
  \begin{align*}
  r_i=\left\{\begin{array}{ll}
  h_{\frac{1}{x}} & \text{if $i=1$}\\
  h_{t} & \text{if $i=2,3,\ldots,\alpha+1$},
  \end{array}\right.
  \end{align*}
    then the the left side of (\ref{eqmehta1}) coincides with the multidimensional integral of our interest in (\ref{mehtaourint}). Under the above parameter selection, however, the right side of (\ref{eqmehta1}) takes the indeterminate form $0/0$. Therefore, we have to evaluate following limit: 
    \begin{equation}\label{khatrilimitq}
	\mathcal{R}_{m}(\beta,\alpha,x,t) = \tilde{\mathcal{C}}_{(0,\beta,m)}\;\lim_{\substack{r_{1}\to h_{\frac{1}{x}}\\r_{2}, r_{3} ,...,r_{\alpha+1}\to h_{t}}}\frac{\det\left[P_{m+i-1}^{(0,\beta)}(r_{j})\right]_{i,j=1,2,...,\alpha+1}}{\Delta_{\alpha+1}(\textbf{r})}.
	\end{equation}
    
 To this end, following Khatri \cite{Khatri}, we have
 \begin{align}\label{eqkhatrilimit}
		&\lim_{\substack{r_{1}\to h_{\frac{1}{x}}\\r_{2}, r_{3} ,...,r_{\alpha+1}\to h_{t}}}\frac{\det\left[P_{m+i-1}^{(0,\beta)}(r_{j})\right]_{i,j=1,2,...,\alpha+1}}{\Delta_{\alpha+1}(\textbf{r})}
        = 
        \frac{\det\left[P_{m+i-1}^{(0,\beta)}(h_{\frac{1}{x}})\hspace{6mm} \frac{{\rm d}^{j-2}}{{\rm d}h_{t}^{j-2}}P_{m+i-1}^{(0,\beta)}(h_{t})\right]_{\substack{i=1,2,...,\alpha+1\\j=2,3,...,\alpha+1}}}{\det\left[h_{\frac{1}{x}}^{i-1}\hspace{6mm} \frac{{\rm d}^{j-2}}{{\rm d}h_{t}^{j-2}}h_{t}^{i-1}\right]_{\substack{i=1,2,...,\alpha+1\\j=2,3,...,\alpha+1}}}
	\end{align}
    Now the determinant in the denominator of (\ref{eqkhatrilimit}) simplifies as
    \begin{align*}
    \det\left[h_{\frac{1}{x}}^{i-1}\hspace{6mm} \frac{{\rm d}^{j-2}}{{\rm d}h_{t}^{j-2}}h_{t}^{i-1}\right]_{\substack{i=1,2,...,\alpha+1\\j=2,3,...,\alpha+1}}=\prod_{j=1}^{\alpha-1}j!(h_{t}-h_{\frac{1}{x}})^{\alpha}.
    \end{align*}
    The numerator can be rewritten with the help of (\ref{jacobiDerivative}) as
    \begin{align*}
    &\det\left[P_{m+i-1}^{(0,\beta)}(h_{\frac{1}{x}})\hspace{6mm} \frac{{\rm d}^{j-2}}{{\rm d}h_{t}^{j-2}}P_{m+i-1}^{(0,\beta)}(h_{t})\right]_{\substack{i=1,2,...,\alpha+1\\j=2,3,...,\alpha+1}}\\
    &\qquad \qquad \quad =2^{-\frac{\alpha}{2}(\alpha-1)}
    \det\left[P_{m+i-1}^{(0,\beta)}(h_{\frac{1}{x}})\hspace{6mm} (m+\beta+i)_{j-2}P_{m+i-j+1}^{(j-2,\beta+j-2)}(h_{t})\right]_{\substack{i=1,2,...,\alpha+1\\j=2,3,...,\alpha+1}}.
    \end{align*}
  Substituting the above two expression into (\ref{eqkhatrilimit}) and then the result into (\ref{khatrilimitq}) gives
  \begin{dmath*}
		\mathcal{R}_{m}(\beta,\alpha,x,t) = \tilde{\mathcal{C}}_{(0,\beta,m)}\frac{t^{\alpha}}{2^{\frac{\alpha}{2}(\alpha+1)}\prod_{j=1}^{\alpha-1}(j)!(1-xt)^{\alpha}}\times \det\left[P_{m+i-1}^{(0,\beta)}\left(h_{\frac{1}{x}}\right)\hspace{6mm} (m+i+\beta)_{j-2}P_{m+i-j+1}^{(j-2,\beta+j-2)}(h_{t})\right]_{\substack{i=1,2,...,\alpha+1\\j=2,3,...,\alpha+1}}.
	\end{dmath*}

	\section{Proof of the microscopic limit of the c.d.f. of the maximum eigenvalue}\label{appb}
    
    Let us rewrite \eqref{cdfXmaxFinite}, keeping in mind  $ \alpha=n-m $ , $ \beta = p-m $, and $ \gamma = m+\alpha+\beta+1 $, as 
        	\begin{dmath}
            \label{asyseed}
		\Pr(x_{\max} \leq t)=t^{m(\alpha+\beta+m)}\left(\prod_{j=0}^{\alpha-1}\dfrac{(\beta+2m+j-1)!}{(\beta+2m+2j)!}\right)\times
		\det\left[\mathcal{P}_{i}(m,\alpha,\beta,\eta,t) \hspace{6mm}(m+i+\beta-1)_{j-2}P_{m+i-j}^{(j-2,\beta+j-2)}\left(\frac{2}{t}-1\right)\right]_{\substack{i=1,2,...,\alpha+1\\j=2,3,...,\alpha+1}}
	\end{dmath}
	where
	\begin{dmath}
		\mathcal{P}_{i}(m,\alpha,\beta,\eta,t)=\frac{\Gamma(\alpha+\beta+2m+i-1)\Gamma(\beta+m+i-1)}{\Gamma(\beta+2m+2i-2)\Gamma(m+\beta)(1+\eta)^{m+\beta}}\left(\frac{\eta t}{1+\eta}\right)^{i-1}
		\times\Hypergeometric{2}{1}{\beta+m+i-1,\alpha+\beta+2m+i-1}{\beta+2m+2i-2}{\frac{\eta t}{1+\eta}}.
	\end{dmath}
	 Following (\ref{jacobidef2}), the Jacobi polynomial $ P_{m+i-j}^{(j-2,\beta+j-2)} $ can be written as
	\begin{align}\label{jacobi1} P_{m+i-j}^{(j-2,\beta+j-2)}\left(\frac{2}{t}-1\right)&=\frac{(j-1)_{m+i-j}}{(m+i-j)!}\Hypergeometric{2}{1}{-(m+i-j),m+\beta+i+j-3}{j-1}{1-\frac{1}{t}},
    \end{align}
    from which we obtain
    \begin{align}
		&P_{m+i-j}^{(j-2,\beta+j-2)}\left(\frac{2}{t}-1\right)\nonumber\\
        &\qquad =\frac{(m+i-2)!}{(m+i-j)!(j-2)!}\sum_{k_{j}=0}^{m+i-j}\frac{(-(m+i-j))_{k_{j}}(m+\beta+i+j-3)_{k_{j}}}{(k_{j})!(j-1)_{k_{j}}}\left(1-\frac{1}{t}\right)^{k_{j}}.
	\end{align}
	To facilitate further analysis, we need to eliminate the dependence of summation upper limit on $i$.  To this end, we decompose the two Pochhammer symbols in the numerator of the above summation as
    \begin{dmath}
		(-(m+i-j))_{k_{j}}=\frac{(-(m+\alpha-j+1))_{k_{j}} (m+i-j-k_{j}+1)_{\alpha-i+1}(m+i-j)!}{(m+\alpha-j+1)!}
	\end{dmath}
	and
	\begin{dmath}
		(m+\beta+i+j-3)_{k_{j}}=\frac{(m+\beta+j-2))_{k_{j}}(m+\beta+j+k_{j}-2)_{i-1} (m+\beta+j-3)!}{(m+\beta+i+j-4)!}.
	\end{dmath}
	Therefore, we obtain
	\begin{dmath}\label{jaclbi2} (m+i+\beta-1)_{j-2}P_{m+i-j}^{(j-2,\beta+j-2)}\left(\frac{2}{t}-1\right)=\frac{(m+i-2)!(m+\beta+j-3)!(j-2)!(m+i-j)!}{(m+i+\beta-2)!(m+i-j)!(j-2)!(m+\alpha-j+1)!}	\mathcal{S}_{k_{j}}(t)\mathcal{U}_{i,j}(m,\alpha,\beta)
	\end{dmath}
where,
	\begin{dmath}
		\mathcal{S}_{k_{j}}(t)=\sum_{k_{j}=0}^{m+\alpha-j+1}\frac{(-(m+\alpha-j+1))_{k_{j}}(m+\beta+j-2)_{k_{j}}}{(k_{j})!(j+k_{j}-2)!}\left(1-\frac{1}{t}\right)^{k_{j}}
	\end{dmath}
	and
	\begin{dmath}
    \label{Udef}
		\mathcal{U}_{i,j}(m,\alpha,\beta)=(m+\beta+j+k_{j}-2)_{i-1}(m+i-j-k_{j}+1)_{\alpha-i+1}.
	\end{dmath}
	Now we substitute \eqref{jaclbi2} into \eqref{asyseed} with some algebraic manipulation to yield
	\begin{dmath}\label{simp1}
		\Pr(x_{\max} \leq t)=t^{m(\alpha+\beta+m)}\left(\prod_{j=0}^{\alpha-1}\dfrac{\mathcal{S}_{k_{j+2}}(t)(\beta+2m+j-1)!(m+\beta+j-1)!}{(\beta+2m+2j)!(m+\alpha-j-1)!}\right)\\\times
		\det\left[\mathcal{P}_{i}(m,\alpha,\beta,\eta,t)\hspace{6mm} \frac{(m+i-2)!}{(m+\beta+i-2)!}\mathcal{U}_{i,j}(m,\alpha,\beta)\right]_{\substack{i=1,2,...,\alpha+1\\j=2,3,...,\alpha+1}},
	\end{dmath}
	from which we obtain after some rearrangements
	\begin{dmath}
		\Pr(x_{\max} \leq t)=t^{m(\alpha+\beta+m)}\left(\frac{(m-1)!}{(m+\alpha+\beta-1)!}\right)\left(\prod_{j=0}^{\alpha-1}\dfrac{\mathcal{S}_{k_{j+2}}(\beta+2m+j-1)!}{(\beta+2m+2j)!}\right)\\\times
		\det\left[\frac{(m+\beta+i-2)!}{(m+i-2)!}\mathcal{P}_{i}(m,\alpha,\beta,\eta,t)\hspace{6mm} \mathcal{U}_{i,j}(m,\alpha,\beta)\right]_{\substack{i=1,2,...,\alpha+1\\j=2,3,...,\alpha+1}}.
	\end{dmath}
       
     For convenience, let us rewrite the above equation as
     \begin{dmath}\label{{cdf_xmax}_part}
		\Pr(x_{\max} \leq t)=t^{m(\alpha+\beta+m)}\left(\prod_{j=0}^{\alpha-1}\dfrac{\mathcal{S}_{k_{j+2}}(t)(\beta+2m+j-1)!}{(\beta+2m+2j)!}\right)
		\det\left[\mathcal{V}_{i}(m,\alpha,\beta,\eta,t)\hspace{6mm} \mathcal{U}_{i,j}(m,\alpha,\beta)\right]_{\substack{i=1,2,...,\alpha+1\\j=2,3,...,\alpha+1}}
	\end{dmath}
	\noindent where
	\begin{dmath}
		\mathcal{V}_{i}(m,\alpha,\beta,\eta,t)=\frac{(m+\beta+i-2)!(\alpha+\beta+2m+i-2)!(\beta+m+i-2)!(m-1)!}{(m+i-2)!(\beta+2m+2i-3)!(m+\beta-1)!(m+\alpha+\beta-1)!(1+\eta)^{m+\beta}}\left(\frac{\eta t}{1+\eta}\right)^{i-1}\\
		\qquad \times\Hypergeometric{2}{1}{\beta+m+i-1,\alpha+\beta+2m+i-1}{\beta+2m+2i-2}{\frac{\eta t}{1+\eta}}.
	\end{dmath}
    Further manipulation of $\mathcal{V}_{i}(m,\alpha,\beta,\eta,t)$ in its current form is an arduous task due to the presence of the hypergeometric function. To this end, noting that $(\alpha+\beta+2m+i-1)-(\beta+2m+2i-2)=-(\alpha+1-i)$, which is a negative integer, we use the hypergeometric transformation  (\ref{hypertransform}) to arrive at
	  	\begin{dmath}\label{column1simp}
		\mathcal{V}_{i}(m,\alpha,\beta,\eta,t)=\frac{(m+i-1)_\beta}{(m-1)_\beta}\frac{(2m+\beta+2i-2)_{\alpha-i+1}}{(m+\beta+i-1)_{\alpha-i+1}}\left(\frac{(\eta t)^{i-1}}{(1+\eta-\eta t)^{m+\beta+i-1}}\right)\times
		\sum_{\ell=0}^{\alpha-i+1}\dfrac{(\beta+m+i-1)_{\ell}(-(\alpha-i+1))_{\ell}}{(\beta+2m+2i-2)_{\ell}\ell!}\left(\dfrac{\eta t}{\eta t -1 -\eta}\right)^{\ell}.
	\end{dmath}
A careful inspection of (\ref{{cdf_xmax}_part}) reveals that the suitable scaling as $m\to\infty$ would be to consider the scaled $t$ given by	
$ t= 1-\dfrac{x}{m^{2}}$. Consequently, we can write (\ref{{cdf_xmax}_part}) as
		\begin{dmath}
        \label{scaledcdf}
		\Pr\left(x_{\max} \leq 1-\dfrac{x}{m^{2}}\right)=\left(1-\dfrac{x}{m^{2}}\right)^{m(\alpha+\beta+m)}\left(\prod_{j=0}^{\alpha-1}\dfrac{\mathcal{S}_{k_{j+2}}\left(1-\dfrac{x}{m^{2}}\right)(\beta+2m+j-1)!}{(\beta+2m+2j)!}\right)\times
		\det\left[\mathcal{V}_{i}\left(m,\alpha,\beta,\eta,1-\dfrac{x}{m^{2}}\right)\hspace{6mm} \mathcal{U}_{i,j}(m,\alpha,\beta)\right]_{\substack{i=1,2,...,\alpha+1\\j=2,3,...,\alpha+1}}.
	\end{dmath}
Now taking the limits of the both sides of (\ref{scaledcdf})  as $ m\to\infty $ yields
	\begin{dmath}\label{simp2}
		\lim_{m\to\infty}\Pr(x_{\max} \leq 1-\dfrac{x}{m^{2}})=e^{-x}\lim_{m\to\infty}\left(\prod_{j=0}^{\alpha-1}\dfrac{\mathcal{S}_{k_{j+2}}\left(1-\dfrac{x}{m^{2}}\right)(\beta+2m+j-1)!}{(\beta+2m+2j)!}\times
		\det\left[\mathcal{V}_{i}\left(m,\alpha,\beta,\eta,1-\dfrac{x}{m^{2}}\right)\hspace{6mm} \mathcal{U}_{i,j}(m,\alpha,\beta)\right]_{\substack{i=1,2,...,\alpha+1\\j=2,3,...,\alpha+1}}\right)
	\end{dmath}
	Towards taking the limit inside the determinant, let us first consider the $\displaystyle \lim_{m\to\infty}\mathcal{V}_{i}\left(m,\alpha,\beta,\eta,1-\dfrac{x}{m^{2}}\right)$.
To this end, noting that 
	$\displaystyle \lim_{m \to \infty}  \frac{(m+i-1)_\beta}{(m-1)_\beta}=1$ and $\displaystyle \lim_{m \to \infty} \frac{(2m+\beta+2i-2)_{\alpha-i+1}}{(m+\beta+i-1)_{\alpha-i+1}}=2^{\alpha-i+1}$, we may determine the limit of \eqref{column1simp} as
	\begin{align*}
\lim_{m\to\infty}\mathcal{V}_{i}\left(m,\alpha,\beta,\eta,1-\dfrac{x}{m^{2}}\right)=2^{\alpha}\sum_{\ell=0}^{\alpha-i+1}\left(\frac{\eta}{2}\right)^{\ell+i-1}\binom{\alpha-i+1}{\ell}=2^{\alpha}\mathcal{T}_{i}(\eta)
\end{align*}
where $\mathcal{T}_{i}(\eta)=2^{\alpha}\left(\frac{\eta}{2}\right)^{i-1}\left(1+\frac{\eta}{2}\right)^{\alpha-i+1}$

	
    Let us Now consider the other columns of the determinant in \eqref{simp2}. Following (\ref{Udef}), we may rewrite
     $ \mathcal{U}(m,\alpha,\beta) $ as
	\begin{dmath*}
		\mathcal{U}_{i,j}(m,\alpha,\beta)=(m+i-j-k_{j}+1)_{\alpha-i+1}(m+\beta+j+k_{j}-2)_{i-1}
		=\prod_{\ell_{1}=0}^{\alpha-i}{(c_{j}-\ell_1)}\prod_{\ell_{2}=0}^{i-2}{(\Delta_{m}-c_j+ \ell_2)}
	\end{dmath*}
		\noindent where, $ c_{j}=m+\alpha-j-k_{j}+1 $ and $ \Delta_{m} = 2m+\alpha+\beta-1 $. Consequently, the terms in determinant in (\ref{simp2}) can be rearranged as
	\begin{dmath*}
		2^{\alpha}\left|
		\begin{array}{ccccc}
			\mathcal{T}_{1}(\eta)& \prod_{\ell_{1}=0}^{\alpha-1}{(c_{2}-\ell_1)}&\cdots&\prod_{\ell_{1}=0}^{\alpha-1}{(c_{\alpha+1}-\ell_1)}\\
			\mathcal{T}_{2}(\eta)&\prod_{\ell_{1}=0}^{\alpha-2 }{(c_{2}-\ell_1)}(\Delta_m-c_{2})&\cdots&\prod_{\ell_{1}=0}^{\alpha-2}{(c_{\alpha+1}-\ell_1)}(\Delta_m-c_{\alpha+1})\\
			\vdots & \vdots & \ddots &\vdots \\
			\mathcal{T}_{\alpha}(\eta)&(c_{2})\prod_{\ell_{2}=0}^{\alpha-2}{(\Delta_m-c_{2}+\ell_{2})}&\cdots&(c_{\alpha+1})\prod_{\ell_{2}=0}^{\alpha-2}{(\Delta_m-c_{\alpha+1}+\ell_{2})}\\
			\mathcal{T}_{\alpha+1}(\eta)&\prod_{\ell_{2}=0}^{\alpha-1}{(\Delta_m-c_{2}+\ell_{2})}&\cdots&\prod_{\ell_{2}=0}^{\alpha-1}{(\Delta_m-c_{\alpha+1}+\ell_{2})}
		\end{array}
		\right|.
	\end{dmath*}
	Towards making the determinant independent of $ \Delta_m $, we perform the following row operations
	\begin{align*}
    R_{i}\to R_{i}+R_{i-1}, \;\; i=2,3,...,\alpha+1
    \end{align*}
  on each row, starting from the second row, to yield 
   
	\begin{dmath*}
		2^{\alpha}\prod_{\ell=0}^{\alpha-1}\left(\Delta_m-(\alpha-1)+2\ell\right)\times\left|
		\begin{array}{ccccc}
			\mathcal{T}_{1}(\eta)& \prod_{\ell_{1}=0}^{\alpha-1}{(c_{2}-\ell_1)}&\cdots&\prod_{\ell_{1}=0}^{\alpha-1}{(c_{\alpha+1}-\ell_1)}\\
			\frac{\mathcal{T}_{2}(\eta)+\mathcal{T}_{1}(\eta)}{\Delta_m-(\alpha-1)}&\prod_{\ell_{1}=0}^{\alpha-2 }{(c_{2}-\ell_1)}&\cdots&\prod_{\ell_{1}=0}^{\alpha-2}{(c_{\alpha+1}-\ell_1)}\\
			\vdots & \vdots & \ddots &\vdots \\
			\frac{\mathcal{T}_{\alpha}(\eta)+\mathcal{T}_{\alpha-1}(\eta)}{\Delta_m+\alpha-3}&(c_{2})\prod_{\ell_{2}=0}^{\alpha-3}{(\Delta_m-c_{2}+\ell_{2})}&\cdots&(c_{\alpha+1})\prod_{\ell_{2}=0}^{\alpha-3}{(\Delta_m-c_{\alpha+1}+\ell_{2})}\\
			\frac{\mathcal{T}_{\alpha+1}(\eta)+\mathcal{T}_{\alpha}(\eta)}{\Delta_m+\alpha-1}&\prod_{\ell_{2}=0}^{\alpha-2}{(\Delta_m-c_{2}+\ell_{2})}&\cdots&\prod_{\ell_{2}=0}^{\alpha-2}{(\Delta_m-c_{\alpha+1}+\ell_{2})}\\	
		\end{array}
		\right|\\
		=\prod_{\ell=0}^{\alpha-1}\left(\Delta_m-(\alpha-1)+2\ell\right)\times\left|
		\begin{array}{ccccc}
			2^{\alpha}\mathcal{T}_{1}(\eta)& 2^{\alpha}\prod_{\ell_{1}=0}^{\alpha-1}{(c_{2}-\ell_1)}&\cdots&2^{\alpha}\prod_{\ell_{1}=0}^{\alpha-1}{(c_{\alpha+1}-\ell_1)}\\
			\frac{\mathcal{T}_{2}(\eta)+\mathcal{T}_{1}(\eta)}{\Delta_m-(\alpha-1)}&\prod_{\ell_{1}=0}^{\alpha-2 }{(c_{2}-\ell_1)}&\cdots&\prod_{\ell_{1}=0}^{\alpha-2}{(c_{\alpha+1}-\ell_1)}\\
			\vdots & \vdots & \ddots &\vdots \\
			\frac{\mathcal{T}_{\alpha}(\eta)+\mathcal{T}_{\alpha-1}(\eta)}{\Delta_m+\alpha-3}&(c_{2})\prod_{\ell_{2}=0}^{\alpha-3}{(\Delta_m-c_{2}+\ell_{2})}&\cdots&(c_{\alpha+1})\prod_{\ell_{2}=0}^{\alpha-3}{(\Delta-c_{\alpha+1}+\ell_{2})}\\
			\frac{\mathcal{T}_{\alpha+1}(\eta)+\mathcal{T}_{\alpha}(\eta)}{\Delta_m+\alpha-1}&\prod_{\ell_{2}=0}^{\alpha-2}{(\Delta_m-c_{2}+\ell_{2})}&\cdots&\prod_{\ell_{2}=0}^{\alpha-2}{(\Delta_m-c_{\alpha+1}+\ell_{2})}
		\end{array}
		\right|.
	\end{dmath*}
	To facilitate further simplification, noting that 
	\begin{align*} R_{1}\to R_{1}+\sum_{i=2}^{\alpha+1}(-1)^{i-1}\left(R_{i}\times(\Delta_m-(\alpha-1)+2(i-2))\right)\left(2^{\alpha}-\sum_{j=0}^{i-2}\binom{\alpha}{j}\right), \end{align*}
	set the $ 1^{st} $ element of the $ 1^{st} $ column to $ 1 $ and
    \begin{align*}
   \lim_{m\to\infty} \frac{\mathcal{T}_{i}(\eta)+\mathcal{T}_{i-1}(\eta)}{\Delta_m-(\alpha-1)+2(i-2)}=0,\;\;i=2,3,\ldots,\alpha+1,
    \end{align*}
	we apply the row operation $ R_{i}\to R_{i}+R_{i-1} $, for $ i=3,4,...,\alpha+1 $, repeatedly to obtain 
	\begin{dmath*}
    \prod_{j=0}^{\alpha-1}\prod_{\ell_j=0}^{j}(\Delta_m-j+2\ell_j)
        \left|
		\begin{array}{ccccc}
			1& *_{2}&\cdots&*_{\alpha+1}\\
			0&\prod_{\ell_{1}=0}^{\alpha-2 }{(c_{2}-\ell_1)}&\cdots&\prod_{\ell_{1}=0}^{\alpha-2}{(c_{\alpha+1}-\ell_1)}\\
			\vdots & \vdots & \ddots &\vdots \\
			0&(c_{2})&\cdots&(c_{\alpha+1})\\
			0&1&\cdots&1\\	
		\end{array}
		\right|.
	\end{dmath*}
	Here the exact form of the $*$ marked entries are tacitly avoided, since they do not contribute to the determination evaluation. As such, by expanding the determinant using the first column, we have
	\begin{dmath*}
		\prod_{j=0}^{\alpha-1}\prod_{\ell_j=0}^{j}(\Delta_m-j+2
        \ell_j)
		\left|
		\begin{array}{ccccc}
			\prod_{\ell_{1}=0}^{\alpha-2}{(c_{2}-\ell_1)}&\prod_{\ell_{1}=0}^{\alpha-2}{(c_{3}-\ell_1)}&\cdots&\prod_{\ell_{1}=0}^{\alpha-2}{(c_{\alpha+1}-\ell_1)}\\
			\prod_{\ell_{1}=0}^{\alpha-3}{(c_{2}-\ell_1)}&\prod_{\ell_{1}=0}^{\alpha-3}{(c_{3}-\ell_1)}&\cdots&\prod_{\ell_{1}=0}^{\alpha-3}{(c_{\alpha+1}-\ell_1)}\\
			\vdots & \vdots & \ddots &\vdots \\
			c_{2}&c_{3}&\cdots&c_{\alpha+1}\\
			1&1&\cdots&1\\	
		\end{array}
		\right|.
	\end{dmath*}
   The above determinant can be simplified using \cite[Lemma A.1]{me10} to yield
	\begin{dmath*}	
	\prod_{j=0}^{\alpha-1}\prod_{\ell_j=0}^{j}(\Delta_m-j+2\ell_j)\tilde{\Delta}_{\alpha}(\tilde{\textbf{c}})
	\end{dmath*}
    where $ \tilde{\Delta}_{\alpha}(\tilde{\textbf{c}}) =\prod_{1\leq j < i \leq \alpha}\left(\tilde{c}_{i}-\tilde{c}_{j}\right) $ with $ \tilde{\textbf{c}}=\left\{\tilde{c}_{1}(k_{2}),\tilde{c}_{2}(k_{3}),\cdots,\tilde{c}_{\alpha}(k_{\alpha+1})\right\} $ and $ \tilde{c}_{j}(k_{j+1})=j+k_{j+1} $.
Now we substitute the above result into \eqref{simp2} to obtain
\begin{dmath*}
\lim_{m\to\infty}\Pr\left(x_{\max} \leq 1-\dfrac{x}{m^{2}}\right)=
e^{-x}\lim_{m\to\infty}\left(\prod_{j=0}^{\alpha-1}\mathcal{S}_{k_{j+2}}\left(1-\dfrac{x}{m^{2}}\right)\times \left(\prod_{j=0}^{\alpha-1}\dfrac{(\beta+2m+j-1)!}{(\beta+2m+2j)!}\right)\prod_{j=0}^{\alpha-1}\prod_{\ell_j=0}^{j}(\tilde{\Delta}_m-j+2\ell_j)\Delta_{\alpha}(\tilde{\textbf{c}})\right)
		=e^{-x}\lim_{m\to\infty}\left(\prod_{j=0}^{\alpha-1}\mathcal{S}_{k_{j+2}}\left(1-\dfrac{x}{m^{2}}\right)\times \prod_{j=0}^{\alpha-1}\prod_{\ell_j=0}^{j}\frac{(2m+\beta+\alpha-j+2l-1)}{2m+\beta+2j-l}\tilde{\Delta}_{\alpha}(\tilde{\textbf{c}})\right)
		=e^{-x}\lim_{m\to\infty}\left(\prod_{j=0}^{\alpha-1}\mathcal{S}_{k_{j+2}}\left(1-\dfrac{x}{m^{2}}\right)\tilde{\Delta}_{\alpha}(\tilde{\textbf{c}})\right).
	\end{dmath*}
	For notational convenience, the index $j$ is shifted forward by one unit to yield
	\begin{dmath}\label{ef2}
		\lim_{m\to\infty}\Pr\left(x_{\max} \leq 1-\dfrac{x}{m^{2}}\right)
		=e^{-x}\lim_{m\to\infty}\left(\prod_{j=1}^{\alpha}\mathcal{S}_{k_{j}}\left(1-\dfrac{x}{m^{2}}\right)\Delta_{\alpha}({\textbf{c}})\right)
	\end{dmath}
	\noindent where,
	\begin{dmath*}
		\mathcal{S}_{k_{j}}(t)=\sum_{k_{j}=0}^{m+\alpha-j}\frac{(-(m+\alpha-j))_{k_{j}}(m+\beta+j-1)_{k_{j}}}{k_{j}!(j+k_{j}-1)!}\left(1-\frac{1}{t}\right)^{k_{j}}
	\end{dmath*}
	and $ \Delta_{\alpha}({\textbf{c}}) =\prod_{1\leq j < i \leq \alpha}\left({c}_{i}-{c}_{j}\right) $ with $ {\textbf{c}}=\left\{{c}_{1}(k_{1}),{c}_{2}(k_{1}),\cdots,{c}_{\alpha}(k_{\alpha})\right\} $ and $ {c}_{j}(k_{j})=j+k_{j} $. Having noted that $ \Delta_{\alpha}({\textbf{c}})$ is independent of $m$ and 
    \begin{align*}
    \lim_{m\to\infty} \frac{(m+\alpha-j-k_j+1)_{k_j}}{m^{k_j}}=1,\;\;\;
    \lim_{m\to\infty} \frac{(m+\beta+j-1)_{k_j}}{m^{k_j}}=1,
    \end{align*}
  we evaluate the limit of $ \mathcal{S}_{k_{j}}\left(1-\dfrac{x}{m^{2}}\right) $ as
	\begin{dmath*}
		\lim_{m\to\infty}\mathcal{S}_{k_{j}}\left(1-\dfrac{x}{m^{2}}\right)
		=\lim_{m\to\infty} \sum_{k_{j}=0}^{m+\alpha-j}\frac{(-(m+\alpha-j))_{k_{j}}(m+\beta+j-1)_{k_{j}}}{k_{j}!(j+k_{j}-1)!}\left(1-\frac{1}{t}\right)^{k_{j}}=\lim_{m\to\infty} \sum_{k_{j}=0}^{m+\alpha-j}\frac{(m+\alpha-j-k_j+1)_{k_j}}{m^{k_j}}\frac{(m+\beta+j-1)_{k_j}}{m^{k_j}}\frac{x^{k_j}}{k_{j}!(k_{j}+j-1)!}\frac{1}{\left(1-\dfrac{x}{m^{2}}\right)^{k_{j}}}
        = \sum_{k_{j}=0}^{\infty}\frac{x^{k_j}}{k_{j}!(k_{j}+j-1)!}.
	\end{dmath*}
	Therefore, \eqref{ef2} simplifies to
	\begin{dmath*}
		\lim_{m\to\infty}\Pr\left(x_{\max} \leq 1-\dfrac{x}{m^{2}}\right)
		=e^{-x}\sum_{k_{1}=0}^{\infty}\sum_{k_{2}=0}^{\infty}\cdots \sum_{k_{\alpha}=0}^{\infty}\prod_{j=1}^{\alpha}\frac{x^{k_{j}}}{k_{j}!(k_{j}+j-1)!}\Delta_{\alpha}({\textbf{c}}),
	\end{dmath*}
	from which we obtain using \cite[Appendix B]{me17}
	\begin{dmath*}
		\lim_{m\to\infty}\Pr\left(x_{\max} \leq 1-\dfrac{x}{m^{2}}\right)=
		e^{-x}\det\left[\mathcal{I}_{j-i}(2\sqrt{x})\right]_{i,j=1,2,\cdots,\alpha}.
	\end{dmath*}
	The above result implies that,
	\begin{dmath*}
		\lim_{m\to\infty}\Pr(m^2(1-x_\max)\leq x)
		=1-e^{-x}\det\left[\mathcal{I}_{j-i}(2\sqrt{x})\right]_{i,j=1,2,\cdots,\alpha}.
	\end{dmath*}
	\noindent Finally, noting that
	\begin{align*}
	\lim_{m\to\infty}F_{m^2(1-x_\max)}(x) =\lim_{m\to\infty}\Pr\left(m^2(1-x_\max)\leq x\right)=\lim_{m\to\infty} F_{\frac{m^2}{1+\lambda_\max}}(x)=F_{X}(x),
	\end{align*}
    we may use the continuous mapping theorem \cite{vaart1998cambridge} to obtain \eqref{thmasylambdamax}, which concludes the proof.

	
	%

	\appendices
	



	\ifCLASSOPTIONcaptionsoff
	\newpage
	\fi

\end{document}